\documentclass[11pt]{article}
\usepackage{amsfonts}
\usepackage{epsfig}

\catcode`\@=11 
\def\numberbysection{\@addtoreset{equation}{section} 
        \def\theequation{\thesection.\arabic{equation}}}

\numberbysection 
 
\newcommand{\beq}{\begin{equation}} 
\newcommand{\eeq}{\end{equation}} 
\newcommand{\bea}{\begin{eqnarray}} 
\newcommand{\eea}{\end{eqnarray}} 
\newcommand{\beas}{\begin{eqnarray*}} 
\newcommand{\eeas}{\end{eqnarray*}} 
 
\def\de{\partial} 
\def\s{\sigma} 
\def\a{\alpha} 
\def\b{\beta} 
\def\d{\delta}
\def\D{\Delta} 
\def\eps{\epsilon} 
 
\def\G{\Gamma} 
\def\l{\lambda}
\def\ex{{\rm e}} 
\def\di{\rm dim}

\def\o{\omega} 
\def\T{\Theta} 
\def\Im{\hbox{\rm Im\ }} 
 
\def\half{{1 \over 2}} 
\def\nl{\nonumber\\} 
\def\bh{\!\!\!\!\!\!} 
\def\TTT{\bra TTT\ket} 
\def\TT{\bra TT\ket} 
\def\p{{\cal P}}
 
\def\bra{\langle} 
\def\ket{\rangle} 
\newcommand{\ndt}[2]{ {\bf #1}_{#2} } 
\newcommand{\ntp}[1]{{N_p(#1)}} 
\newcommand{\nta}[1]{{N(#1)}}   
\newcommand{\ntb}[1]{{N_b(#1)}} 
 
\def\dt{{\delta_T}} 
\def\mup{{\mu'}}    
\def\gflat{{\eta}}  
\def\A{{\cal A}}    
\begin{document}  
 
 
\begin{titlepage} 
\begin{center} 
 
\hfill SPhT T-01/027 \\
\hfill  DFF 371/01/2001 \\ 
\hfill  GE-TH-03/2001\\ 
\hfill hep-th/0103237\\
 
\vspace{1.cm} 
 
{\bf\Large Exact Consequences of the Trace Anomaly} \\ 
 
\bigskip 
{\bf\Large in Four Dimensions}\\ 
 
\vspace{1cm} 
 
Andrea CAPPELLI\\ 
\normalsize\textit{ I.N.F.N. and Dipartimento di Fisica,} \\ 
\normalsize\textit{ Largo E. Fermi 2, I-50125 Firenze, Italy}\\ 
\normalsize\textit{ e-mail: andrea.cappelli@fi.infn.it}\\ 
 
\vspace{.3cm} 
 
Riccardo GUIDA\\ 
\normalsize\textit{ CEA-Saclay, Service de Physique Th\'eorique,} \\ 
\normalsize\textit{ F-91191 Gif-sur-Yvette, France}\\ 
\normalsize\textit{ e-mail: guida@spht.saclay.cea.fr}\\ 
 
\vspace{.3cm} 
 
Nicodemo MAGNOLI\\ 
\normalsize\textit{ I.N.F.N. and Dipartimento di Fisica,} \\ 
\normalsize\textit{ Via Dodecaneso 33, I-16146 Genova, Italy}\\ 
\normalsize\textit{ e-mail: magnoli@ge.infn.it.}\\ 
 
\vspace{.5cm} 
 
\begin{abstract} 
The general form of the stress-tensor three-point function in four  
dimensions is obtained by solving the Ward identities 
for the diffeomorphism and Weyl symmetries.
Several properties of this correlator are discussed, such as
the renormalization and scheme independence and the analogies
with the anomalous chiral triangle.
At the critical point, the coefficients $a$ and $c$ of the four-dimensional 
trace anomaly are related to two finite, scheme-independent
amplitudes of the three-point function.
Off-criticality, the imaginary parts of these amplitudes satisfy sum rules
which express the total renormalization-group flow 
of $a$ and $c$ between pairs of critical points.
Although these sum rules are similar to that satisfied by the two-dimensional 
central charge, the monotonicity of the flow, i.e.
the four-dimensional analogue of the $c$-theorem, remains to be proven. 
\end{abstract} 
 
\vfill 
\hfill February 2001
\end{center} 
 
\end{titlepage} 
\pagenumbering{arabic}


\section{Introduction}

The Zamolodchikov $c$-theorem \cite{cth} states that the
two-dimensional unitary field theories are characterized by a 
positive function of the coupling constants $c(g)$ which
decreases along the renormalization-group flow;
moreover, its value at the fixed points is equal 
to the trace-anomaly number $c$.
Several works have suggested that the theorem could extend to 
four dimensions\footnote{
For a short review, see Ref.\cite{cdgm}; for early attempts, see the
Refs.\cite{cardy}\cite{jo}\cite{cfl}.}:
at criticality, the candidate $c$-function would again reduce 
to the coefficient of the the Euler density in the trace anomaly \cite{cardy}.
This conjecture is supported by many examples 
of non-trivial flows in supersymmetric gauge theories,
which have been analysed {\it via} the Seiberg-Witten  
electric-magnetic duality \cite{afgj} 
and the AdS supergravity-gauge theory correspondence \cite{fgpw}\cite{c-a}. 
Two proofs of the four-dimensional theorem
have also been proposed, although one is rather controversial
\cite{fl}, and the other one relies on a strong dynamical hypothesis
\cite{ath}.

These results and conjectures motivate further analyses of the physical
consequences of the four-dimensional trace anomaly\footnote{ 
Note that the normalization of the four-dimensional
stress tensor in Eq.(\ref{theta-def}) 
differs from that in Refs.\cite{cfl}\cite{ad} by the factor
$1/4\pi^2$; nevertheless, the values of the anomaly coefficients 
for the free conformal-invariant scalar field are still normalized to 
$a=c=1$.}:
\beq\label{theta-def} 
g^{\mu\nu} \bra T_{\mu\nu} \ket \equiv 
\bra \Theta \ket = \lambda \left(
a\ E -3c\ W +a'\ D^2 {\cal R}+ r\ {\cal R}^2 \right)\ , \qquad
\lambda\equiv \frac{1}{2880\cdot 4\pi^2}\ .
\eeq  
In this equation, $E$ denotes the Euler density 
($\chi\propto\int \sqrt{g}\ E$ is the
Euler characteristic), $W$ the square of the Weyl tensor and ${\cal R}$  the 
curvature scalar (see Appendix \ref{app-conv} for our conventions). 
Some properties of the trace anomaly (\ref{theta-def}) are known  
since a long time \cite{chi-book}\cite{duff}: the fact that the  
Wess-Zumino consistency conditions imply $r=0$ \cite{pbb}\cite{cc} and 
that the coefficient $a'$ is scheme dependent, because it can be changed 
by adding a local counterterm to the effective action \cite{duff}. 
An interesting study of the four-dimensional trace anomaly 
(and its higher-dimensional analogues) has been done in Ref.\cite{ds}, 
by deriving the corresponding terms in the regularized effective action
within the framework of the epsilon-expansion.
It was found that the topological invariant Euler term
corresponds to an effective action which is finite and 
non-local as $\eps\to 0$; this is the same feature of the chiral anomaly 
\cite{chi-book}.  
On the other hand, the $W$ term in (\ref{theta-def}) 
corresponds to a singular effective action with 
finite scale variation.
 
In this work, we relate the four-dimensional trace anomaly to
some physical quantities:
{\it i)} we express the universal parameters $a$ and $c$ in 
terms of {\it finite} scheme-independent correlation functions 
at criticality; {\it ii)} away from criticality, we provide
{\it exact} sum rules for the renormalization-group flow\footnote{
These flows have also been analysed by the Refs.\cite{jo}\cite{ath} 
within the perturbative epsilon-expansion.}  
of $a$ and $c$.

The analogous two-dimensional results are
well known \cite{pdf}: the coefficient $c$ of the trace anomaly
\hbox{$\bra \Theta\ket = -c \ {\cal R}/ 12$} 
is unambiguously defined by the stress-tensor two-point function  
$\bra T(z) T(0)\ket= c/2 z^4$ at criticality, which
is a finite and well-defined correlator.
Furthermore, the renormalization-group flow of $c$ is 
constrained by the Cardy sum rule, which involves the off-critical 
correlator \cite{sumrule}:
\beq
c_{UV}-c_{IR}= \frac{3}{4\pi}\int_{|x|>\eps}\ d^2x\ x^2 \bra
\Theta(x)\Theta(0)\ket = \int_\eps^\infty\ d s\ c(s) \ >\ 0\ .
\label{2d-sum}
\eeq
In the above Equation, $c_{UV}$ and $c_{IR}$ are the central charges
of, respectively, the ultraviolet and infrared fixed points
connected by the renormalization-group trajectory,
and the $x$-integral is evaluated at any point along the trajectory.
Equation (\ref{2d-sum}) also shows the form of the sum rule
in momentum space \cite{cfl}: it involves 
the {\it positive-definite} spectral measure $c(s)\ ds$, 
which is obtained from the imaginary part of $\bra\Theta(p) \Theta(-p)\ket$
at momentum $p^2=s$.
In four dimensions, the relation of the trace anomaly (\ref{theta-def}) 
to finite correlators requires the study of  
the three-point function in flat space\footnote{
Hereafter, we shall mostly use the Euclidean signature
(see Appendix A for our conventions).}
or of the two-point function in curved space;
their general expressions at criticality (in coordinate space) 
have been found in the Refs.\cite{eo}\cite{ol} and \cite{os}, respectively. 

In this paper, we obtain the general form of the three-point 
function in (flat) momentum space {\it at} and {\it off} criticality:
we write it as a sum of scalar amplitudes (form factors) times
tensor structures, which solve the Ward identities for the diffeomorphism 
and Weyl symmetries (the latter being imposed at criticality).
In addition, we write dispersion relations for the renormalized amplitudes 
following closely the well-known analysis 
of the chiral triangle \cite{fsby}. 

At criticality, we can identify the three-point amplitudes
which correspond to the terms in the trace anomaly,
disentangling the scheme-dependent amplitude 
related to $a'$ in (\ref{theta-def}) from the scheme-independent ones
corresponding to the coefficients $a$ and $c$. 
Any amplitude can be singled out from the general expansion of 
$\TTT$ by solving a linear system:
we obtain some {\it reduction formulae} which can be useful 
for computing $a$ and $c$ in interacting theories;
these relations could also suggest further physical consequences
of the trace anomaly.

Away from criticality, the imaginary parts of the $a$ and $c$
amplitudes satisfy two exact sum rules, which express the total 
variation of $a$ and $c$ under the renormalization-group flow 
between two fixed points; these sum rules follow from the analysis
of the general structure of $\TTT$ 
and are analogous of the two-dimensional result (\ref{2d-sum}).

In two dimensions, the imaginary part of
the two-point function $c(s)$ is positive definite
by unitarity; the sum rule (\ref{2d-sum}) then
implies that $c$ decreases along the renormalization-group flow.
In four dimensions, the property $a_{UV} > a_{IR}$ could be
similarly proved if the imaginary part of the $a$ amplitude
were positive definite.
In general, we do not expect that three-point functions,
related to decay amplitudes, should have a definite sign;
nevertheless, we should mention the positivity condition proposed in 
Ref.\cite{ol}, coming from the classical weak-energy 
condition \cite{bd}, which implies the positivity of the $a$ 
coefficient {\it at criticality}.
The consequences of this condition on the
three-point amplitudes off-criticality remain to be analysed.

The plan of the paper is the following: in Section 2, we briefly 
recall the dispersive analysis of the chiral triangle 
$\bra AVV \ket$ of Ref.\cite{fsby}; this lets us
outline some features which arise for $\bra TTT\ket$ too. 
In this Section, we also give the definition
of the stress-tensor operator and repeat
the dispersive analysis of $\bra TT \ket$ \cite{cfl}\cite{clv}\cite{hore}.
Appendix \ref{app-conv} and \ref{app-ward} contain our notations
for the curved-space calculus and the derivation of the 
Ward identities, respectively. 

In Section 3, we present the analysis of $\bra TTT \ket$: 
{\it i)} the decomposition in scalar amplitudes times the tensors
 solving the diffeomorphism Ward identity
(the tensor basis is constructed in Appendix \ref{app-tens}); 
{\it ii)} the conditions on the amplitudes 
which are imposed by the (anomalous) Weyl symmetry at criticality.
Owing to the large number of possible tensor structures,
we make an extensive use of algebraic programs
(longer formulas are given in Appendix \ref{app-sol} and the full output
is available as a Mathematica \cite{wolf} notebook). 
In this Section, we also discuss the Ward identities that
follow from flat-space conformal invariance at criticality.

In Section 4, we show how to extract the scalar amplitudes in
the $\TTT$ expansion from an explicit expressions of this correlator;
in particular, we obtain the reduction formulae which determine
the trace anomaly coefficients $a$ and $c$ at criticality. 
Next, we discuss the renormalization-group flow of the corresponding
amplitudes off-criticality and derive the sum rules analogues to
Eq.(\ref{2d-sum}).
In Section 5, we test the reduction formulas and the sum rules
by computing the flows of the free massive scalar and fermion theories.
In the Conclusions, we briefly discuss the possible
attempts towards the $c$-theorem in four dimensions.


\section{Dispersive analysis of the chiral triangle 
and of the stress-tensor two-point function}


\subsection{Analysis of $\bra AVV \ket$} 
 
We start by recalling the well-known analysis 
of the three-point function of one axial and two vector currents \cite{fsby}: 
\bea\label{G-def} 
\G_{\mu\a_1\a_2} &\equiv & \int \ d^4x_1 d^4 x_2 \  
\ex^{-ik_1 x_1 -i k_2 x_2} \  
\bra A_\mu (0) V_{\a_1} (x_1) V_{\a_2} (x_2) \ket\ ,\nl 
q^\mu &=& k_1^\mu +k_2^\mu\ , \qquad\quad  
k_1^2=k_2^2\equiv k^2 \ . 
\eea 
We shall discuss the {\it renormalized} correlator,
i.e. we shall not explicitly introduce bare expressions and counterterms,
although some finite scheme-dependent terms will occur.
The space-time symmetries allow to decompose (\ref{G-def}) 
in tensor structures times scalar amplitudes 
$A_i=A_i(q^2,k_1^2,k_2^2)$; we consider the symmetric point 
$k_1^2=k_2^2$, in order to fully exploit the Bose symmetry  
$\left( k_1^{\a_1} \leftrightarrow k_2^{\a_2} \right)$ of the correlator.
The tensor notation can be simplified  
by contracting the Lorentz indices with general polarization 
vectors $\left\{v^\mu, e_1^\mu,e_2^\mu \right\}$, following 
Ref.\cite{rose}. We then write: 
\bea\label{G-exp} 
\G\left(v,e_1,e_2\right) &\equiv&  
v^\mu\ e_1^{\a_1}\ e_2^{\a_2}\ \G_{\mu\a_1\a_2} =  
  A_1 \ \left[ v e_1 e_2 (k_1-k_2) \right]  \nl 
& +& A_3 \left( \left[ v e_1 k_1 k_2 \right] (e_2.k_1) +  
                \left[ v e_2 k_2 k_1 \right] (e_1.k_2) \right) \nl  
& +& A_5 \left( \left[ v e_1 k_1 k_2 \right] (e_2.k_2) +  
                \left[ v e_2 k_2 k_1 \right] (e_1.k_1) \right) \nl 
& +& A_7 \left[e_1 e_2 k_1 k_2 \right] \left( v.(k_1+k_2) \right)\ , 
\eea  
where we denoted: 
\beq 
a_\mu b^\mu \equiv (a.b) \ ,\qquad\quad 
\eps_{\mu\nu\rho\s} a^\mu b^\nu c^\rho d^\s \equiv  
\left[ abcd \right] \ . 
\eeq 
The tensor structures in (\ref{G-exp}) are not linearly independent  
in four dimensions, due to the identity \cite{rose}: 
\beq\label{rose-ide} 
(a.f)\left[bcde\right] + (b.f)\left[cdea\right] +  
(c.f)\left[deab\right] + (d.f)\left[eabc\right] +  
(e.f)\left[abcd\right] =0\ , 
\eeq 
which holds for any six-plet of vectors $\left\{a,b,c,d,e,f\right\}$. 
In its Bose symmetrized form, this identity implies  
that one of the scalar amplitudes in the expansion (\ref{G-exp}) 
is redundant. Actually, $\G$ is invariant under 
the shift of amplitudes by an arbitrary function $f$: 
\beq\label{shift} 
A_1 \to A_1 - \left( k^2 +(k_1.k_2) \right)f \ ,\qquad\quad 
A_i \to A_i - f \ ,\qquad {\rm for}\ \ i=3,5,7\ . 
\eeq 
Therefore, we can eliminate one of the amplitudes in (\ref{G-exp}): 
we choose $A_7=0$ to match the notations of Ref.\cite{fsby}. 
 
We consider a general theory with massive fermions and 
write the Ward identities for the conservation of the vector and
axial currents which enter in $\bra AVV \ket$ 
(the axial current is supposed anomalous).
These Ward identities can be expressed in our notation by choosing
longitudinal polarizations in $\G\left(v,e_1,e_2 \right)$, for example
$e_1= k_1$. We then write,
\bea
\G\left(v,k_1,e_2 \right) &=& 0 \ ,\nl
\G\left(k_1+k_2,e_1,e_2 \right) &=& 
2m \ \G_5\left(e_1,e_2 \right)\ +\ 
{C \over \pi}\left[e_1 e_2 k_1 k_2 \right]\ ;
\label{chiwi}
\eea
$\G_5\left(e_1,e_2 \right)$ is the Fourier transform of the
corresponding triangle $\bra J_5 V V \ket$, involving
the pseudo-scalar density, whose tensor structure is:
\beq
\G_5\left(e_1,e_2 \right) = -\ A_9\ \left[e_1 e_2 k_1 k_2 \right]\ .
\label{G5}
\eeq
The resulting conditions on the scalar amplitudes are: 
\bea 
2 A_1 &=& q^2 A_3 + 2k^2 \left(A_5-A_3 \right)\ ;\label{v-cons}\\ 
2 A_1 &=&  -{ C\over \pi} + 2 m\ A_9\ . 
\label{a-cons} 
\eea 

The quantity $C$ on the right-hand side of (\ref{a-cons}) is the  
chiral anomaly: $C$ cannot be put to zero 
because $\G$ would vanish altogether for $m=0$ at the point $q^2=k^2=0$, 
which is not physically acceptable.
Furthermore, $C$ can be shown to be a constant as follows \cite{fsby}.
Consider the imaginary parts of all amplitudes w.r.t. both variables  
$q^2$ and $k^2$: they are finite quantities  
describing physical scattering processes, that satisfy 
non-anomalous Ward identities,
\bea 
2 \ \Im A_1(q^2,k^2) &=& \left( q^2 -2 k^2 \right) \Im A_3(q^2,k^2)+
2 k^2\ \Im A_5 (q^2,k^2)\ , 
\label{im-v-cons} \\ 
\Im A_1(q^2,k^2) &=& m\ \Im A_9(q^2,k^2) \ .
\label{im-a-cons}
\eea 
It follows that the anomalous term has vanishing imaginary parts:
\beq
C(q^2+i\eps,k^2)-C(q^2-i\eps,k^2) =0=
C(q^2,k^2+i\eps)-C(q^2,k^2-i\eps);
\eeq
thus, $C$ is a dimensionless polynomial of $q^2/m^2$ and $k^2/m^2$ with
finite $m\to 0$ limit, i.e. it is a constant.
 
Next, we write the dispersion relations w.r.t. $q^2$: 
\bea\label{int-a1} 
&&A_1(q^2,k^2)=A^{UV}_1 +{1\over \pi }\int ds\  
{\Im A_1(s,k^2) \over s- q^2} \ , 
\\ 
\label{int-a3} 
&& A_i(q^2,k^2) = {1\over \pi }\int ds\  
{\Im A_i(s,k^2) \over s- q^2} \ , \quad\qquad i=3,5,9.
\eea 
The integrals in Eq.(\ref{int-a1},\ref{int-a3}) 
are convergent due to the asymptotic behaviors,
\beq 
A_i(q^2,k^2)\sim q^{\ \di A_i }\ , \qquad {\rm for}\; q^2\to \infty \ ,
\eeq
following from dimensional analysis (up to logarithms which are supposed
absent for $A_1$):
$\di A_1 = 0$, $\di A_3 =\di A_5 =-2$ and $\di A_9 =-1$; 
only the amplitude $A_1$ (superficially divergent at the bare level)
requires the subtraction term $A_1^{UV}$ in (\ref{int-a1}). 
Combining Eqs. (\ref{im-v-cons},\ref{im-a-cons}) and  
(\ref{int-a3}), we can derive the sum rule: 
\beq\label{sumrule} 
{1\over \pi }\int ds\ {\Im A_3(s,k^2) } + 
\left( q^2-2 k^2 \right) A_3(q^2,k^2) + 2 k^2 A_5(q^2,k^2) =
2 m A_9 (q^2,k^2) \ . 
\eeq 

The comparison of this sum rule with the original Ward identities
(\ref{v-cons},\ref{a-cons}) determines $A_1^{UV}$
in (\ref{int-a1}) as the chiral anomaly $C$ and yields the following
sum rule:
\beq\label{a-const} 
A_1^{UV} = - {C \over 2\pi} = - {1 \over 2\pi} \int \ ds\  
\Im A_3(s,k^2)\ ,   
\eeq 

Let us further analyse the Eqs.
(\ref{v-cons},\ref{a-cons},\ref{a-const}) for $k^2=m^2=0$:
$A_5$ and $A_9$ drop out and we find the explicit solution,
\bea\label{chi-sol} 
A_3(q^2,0) &=& -{ C \over \pi}{1\over q^2} \ , \qquad\quad  
\Im A_3(q^2,0)= C \delta(q^2)\ , \nl 
A_1(q^2,0) &=& - {C \over 2\pi} = - {1 \over 2\pi} \int \ ds\  
\Im A_3(s,0) \ . 
\eea 
This is famous result \cite{fsby} that the anomaly appears as
a delta term in the imaginary part of a finite amplitude 
($A_3$).
We also notice that in the general case
$k^2\neq 0, m \neq 0$, the constancy of the sum rule (\ref{a-const})
implies a non-trivial constraint on $\Im A_3(q^2,k^2,m)$:
this should be a {\it smoothed delta-function }
with spreading parameter proportional to $k^2$ for $m^2=0$
(respectively, $m^2$ for $k^2=0$). 

Let us summarize a number of facts which will  
also occur in the analysis of $\bra TTT \ket$: 
\begin{itemize} 
\item  
There could be algebraic degeneracies among the tensor structures  
which can be written for the three-point function. 
\item 
The anomaly can be expressed in terms of  
a finite, scheme-independent amplitude of the three-point function  
(i.e. $ A_3(q^2,0) \propto C/q^2$); this relation is analogous to 
$\bra T(z)T(0)\ket =c/2z^4$ in two dimensions.
\item 
Scheme-dependent amplitudes (i.e. $A_1$) are related to scheme-independent,
finite amplitudes (i.e. $A_3$) by the Ward identities. 
\item
The sum rule for the chiral anomaly (\ref{a-const}) is similar 
to the sum rule (\ref{2d-sum}) for the conformal anomaly in two dimensions,
which is a version of the $c$-theorem: both expressions
involve dimensionless densities which are smoothed delta-functions.
This similarity will be fully explained
in Section 4, where the sum rules for the four-dimensional
trace anomalies will be derived by combining both arguments.
\end{itemize} 

 
\subsection{Definition of the stress tensor and Ward identities} 

Here we consider a four-dimensional
renormalizable field theory and suppose that it
can be covariantly extended on classical curved-space backgrounds;
at critical points, we also assume that the theory is covariant under 
Weyl transformations of the background metric  $g_{\mu\nu}(x)$.
 The stress-tensor $n$-point function is defined by the following 
variation of the generating functional $W[g]\equiv\log Z[g]$ 
w.r.t. the metric: 
\beq\label{tn-def} 
\bra T_{\mu_1 \mup_1}(x_1) \cdots T_{\mu_n \mup_n} (x_n)\ket \equiv  
{1 \over \sqrt{g(x_1)}\cdots \sqrt{g(x_n)}}  
\ {\delta\over \delta g^{\mu_1 \mup_1}(x_1)} \cdots 
{\delta\over \delta g^{\mu_n \mup_n}(x_n)} W[g] \ . 
\eeq 
In this Equation, the $\sqrt{g}$ terms are pulled out of 
the variations, so that $n$-point functions {\it explicitly} respect  
the symmetry under exchanges of two stress tensors. 
Other definitions would yield $n$-point functions differing 
by lower $k$-point functions, $0\le k <n$; this is an intrinsic 
ambiguity of the stress-tensor which should be fixed together with
the renormalization conditions
(this point will be further discussed in Section 5). 

The symmetries of coordinate reparametrizations
(Diff) and, at critical points, of Weyl rescalings of the metric 
imply Ward identities which can be obtained by varying $W[g]$ and by
consistently using the definition of the stress tensor (\ref{tn-def})
(see Appendix \ref{app-conv} for the conventions of curved-space calculus 
and Appendix \ref{app-ward} for a complete discussion of the Ward identities).
We can write:
\bea\label{ward-def}
\bh {\rm Diff}:\  \d_\eps W[g] &\equiv& 
\int dx\ \left(D^{\mu} \eps^{\nu}(x)+ D^{\nu} \eps^{\mu}(x) \right)
{\d \over \d g^{\mu\nu}(x)} W[g]=0 \ ; \nl 
\bh {\rm Weyl}:\ \d_\s W[g] &\equiv& 
- \int dx\ 2 \s (x)\ g^{\mu\nu}(x) {\d \over \d g^{\mu\nu}(x)} W[g] \nl
&=&- \int dx\ 2 \s(x)\ \sqrt{g}\ {\cal A}(x;g)  \ . 
\eea 
Note that the Weyl symmetry is anomalous with trace anomaly ${\cal A}(x;g)$.
The first variation w.r.t. $g_{\mu\nu}$ of the identities
(\ref{ward-def}) yield the Equations: 
\bea 
\bh\bh {\rm Diff}: &&  D^{\mu} \bra T_{\mu \mup}\ket = 0\ ; \label{1p-diff}\\ 
\bh\bh {\rm Weyl}: &&  g^{\mu \mup}\bra T_{\mu \mup}\ket \equiv  
\bra \Theta\ket = \A(x;g) =\l\left(  
a\ E -3c\ W +a'\ D^2 {\cal R}+ r\ {\cal R}^2 \right).
\label{1p-weyl} 
\eea 
The Ward identities for multi-point functions  
are obtained by further variations w.r.t. $g^{\mu\mup}(x)$ 
of the Eqs. (\ref{1p-diff})(\ref{1p-weyl}).

The expression (\ref{1p-weyl}) of the trace does not
contain explicit dynamical fields which would violate
the Weyl symmetry at the classical level
$\left( \Theta_{classical}\equiv 0 \right)$:
the underlying hypothesis is that the scale-invariant theory 
in flat space can be extended to a Weyl-invariant theory in curved space.
Above two dimensions, this extension {\it is not automatic}:
flat-space scale invariance is compatible with
derivative fields in the trace, which, however, can often be removed by 
redefining (i.e. ``improving'') the stress tensor \cite{polch}
(obstructions to improvement are discussed, e.g., in Ref.\cite{cfl}).

\subsection{The form of $\bra TT\ket$ and its scheme dependencies}

The general form of the stress-tensor two-point function is 
one of the ingredients which are needed for the construction of the 
three-point function: we recall here the analysis of Ref. \cite{cfl}, which 
can also be found in \cite{duff}\cite{hore}. 
From the variation of (\ref{1p-diff}), we obtain the diffeomorphism Ward
identity for the two-point function in flat space:  
\beq\label{2p-difx} 
2 {\de\over\de y_\mu} \bra T_{\mu\nu}(y)T_{\a\b}(x)\ket + 
{\de\over\de y^{(\a} } \left( \d (y-x) \bra T_{\b)\nu}(x)\ket \right) +
\left( {\de\over\de y^\nu}\d(y-x) \right) \bra T_{\a\b}(x) \ket =0 \ , 
\eeq 
where the symmetrization of indices is 
$x^{(\mu} y^{\nu)}= x^\mu y^\nu + y^\mu x^\nu $.
The standard normalization condition would fix the constant
$\bra T_{\mu\nu}\ket = 0$ in flat space; at critical points, 
this is consistent with the absence of dimensionful parameters in the 
trace anomaly, such as a cosmological constant.
On the other hand, we prefer to keep a non-vanishing one-point
function for the later purpose of studying the scheme-dependence of 
$\bra TTT\ket$ due to off-critical mass parameters;
thus, $\bra T_{\mu\nu}\ket \propto \eta_{\mu\nu}\ m^4$.

The first variation of (\ref{1p-weyl}) similarly yields the
Ward identity for the Weyl symmetry at criticality: 
\beq
\label{2p-weyx} 
\bra \T (y) T_{\a\b}(x) \ket = \l\  
a'\ \left(\eta_{\a\b} \de^2  - \de_\a \de_\b \right) 
\de^2 \d (x-y) \ , \qquad ({\rm critical}\ {\rm points}). 
\eeq 
It is convenient to contract the indices of  
the stress-tensor with generic polarization tensors $h_{\mu\nu}=h_{\nu\mu}$, 
and use the following notations: 
\beq\label{h-def} 
h_\mu^\mu \equiv (h)\ , 
\qquad\quad (h_1)_{\a\b} (h_2)^{\a\b} \equiv (h_1 . h_2)\ ,
\qquad\quad p^\a h_{\a\b} q^\b \equiv (p.h.q)\ .
\eeq 
The conservation and trace equations in momentum space 
will be obtained by respectively choosing the polarizations:
\beq\label{c-t-tens} 
h_i \to \half \left(v \otimes k_i +k_i \otimes v \right)\ , 
\qquad\qquad h_i \to \eta \ ,\qquad\qquad i=1,2, 
\eeq 
where $v$ is an arbitrary four-vector, $k_i$ is the momentum 
of the Fourier transformed tensor $T(k_i)$  
and $\eta$ the Euclidean (Minkowskian) metric
(our conventions are listed in Appendix \ref{app-conv}).
 
In conclusion, the Diff and Weyl
Ward identities for the two-point function in momentum space read:
\bea 
\bh\bh {\rm Diff}:
&&\bh \bra \left(v\otimes p.T(p)\right) \left(h_2.T(-p)\right)\ket +
2 \bra \left( p.h_2.T.v \right) \ket + 
(v.p) \bra \left( h_2.T \right) \ket =0\ ,
\label{2p-diff}\\ 
\bh\bh {\rm Weyl}:
&&\bh \bra \T(p) \left(h_2.T_2(-p)\right) \ket =\l \  a' 
\ p^2 \left( p^2 (h_2)- (p.h_2.p) \right), 
\quad  ({\rm critical}\ {\rm points}).
\label{2p-weyl} 
\eea 
 
The general form of the two-point function in momentum space is obtained
by the procedure of Section 2.1:
the general expansion in tensor structures times scalar 
amplitudes is constrained by the Ward identities.
The one-point function can be written:
\beq
\bra \left( h.T \right)\ket = - 2\ (h)\ f_\Lambda \ ,
\label{1p-t}\eeq
where $f_\Lambda$ is a dimensionful constant related to the mass.
The two-point function solution of the Diff Ward identity
(\ref{2p-diff}) is found to be \cite{cfl}: 
\bea\label{2p-tt} 
\bra \left(h_1.T(p)\right) \bh && \bh \left(h_2.T(-p)\right)\ket \nl 
&=& \left( f_0\left(p^2 \right)-f_2\left(p^2 \right) \right) 
    \left(p^2 (h_1) - (p.h_1.p) \right) 
    \left(p^2 (h_2) - (p.h_2.p) \right) \nl 
&+& (d-1)f_2\left(p^2 \right)\left[(p.h_1.p)(p.h_2.p) -2p^2 (p.h_1.h_2.p) 
               + p^4 (h_1.h_2) \right] \nl 
&+& f_\Lambda\ \left[ (h_1)(h_2) +2 (h_1.h_2) \right] \ . 
\eea 
The two independent scalar amplitudes 
$f_0\left(p^2 \right)$ and $f_2\left(p^2 \right)$
are dimensionless functions of the squared momentum and the mass;
they parameterize the propagation of  
intermediate states of spin zero and two, respectively, and
are multiplied by transverse tensors. 
Moreover, the spin-two tensor is traceless (in two dimensions, 
it vanishes identically providing another example of tensor degeneracy). 
For later purposes, it is convenient to write the  
solution for general space-time dimension $d$, and set $d=4$ when comparing 
with the trace anomaly. 
  
At the critical points ($m=0$), 
the two dimensionless amplitudes can depend
logarithmically on the momentum or be constant:
the spin-zero amplitude is completely determined by 
the Weyl Ward identity (\ref{2p-weyl}):
\beq\label{f0} 
f^c_0\left(p^2 \right) = {\l \ a'\over d-1}\ , \qquad\qquad 
 \qquad\qquad ({\rm critical}\ {\rm points}). 
\eeq 
The absence of a logarithmic term is due to our
starting hypothesis of Weyl invariance at criticality, which forces the  
trace for the stress tensor to vanish as an operator; its correlators
can only contain scheme-dependent contact terms.
Thus, the spin-zero part of $\bra TT\ket$ is a polynomial of
momenta of dimension four which can be modified
by adding counterterms to the effective action.
This verifies the well-known
scheme dependence $a' \to a' + const$ in the trace anomaly (\ref{theta-def})
\cite{duff}.

The amplitude $f^c_2\left(p^2 \right)$ has instead a logarithmic 
singularity at the bare level; its renormalized expression is:
\beq\label{f2} 
f^c_2\left(p^2 \right) = -\l\ c\ \log\left( {p^2\over \mu^2} \right)\ ,
 \qquad\qquad ({\rm critical}\ {\rm points}). 
\eeq 
In the above Equation, $\mu$ is the renormalization scale and 
$c$ is the coefficient of the Weyl term in the trace anomaly 
(\ref{1p-weyl}): 
the latter fact can be proved \cite{duff}\cite{cfl}
by using the Ward identity for scale invariance
in flat space (a conformal isometry fully discussed in Section 3.3).
Equation (\ref{f2}) shows that $f^c_2\left(p^2 \right)$ is defined up to a  
scheme-dependent constant, while the coefficient $c$ is  
scheme-independent: its value normalizes the two-point function at
criticality (for non-coincident coordinates, $x_1\neq x_2$) 
and is positive for unitary theories.

In conclusion, the two $\bra TT\ket$ amplitudes depend on the renormalization
scheme by constant shifts, which change the correlator by
{\it polynomials} of the momenta.
The scheme-dependence of any correlator can be characterized as follows:
renormalization theory says that counterterms in the action
are polynomial of momenta with maximal degree 
$d$, thus scheme-dependent terms of a $\D$-dimensional $n$-point function  
$\bra {\cal O}_1(k_1)\cdots {\cal O}_n (k_n) \ket $ are polynomials of
maximal dimension $\D$ ($\D >0$ is assumed here). 
These polynomials can be added to the $n$-point function,
which is being expanded in tensor structures times scalar amplitudes,
as done for $\bra AVV\ket$ and $\bra TT\ket$;
if a polynomial matches one of the tensors in the expansion, 
the corresponding scalar amplitude is redefined.

This scheme dependence can be characterized
by a number associated to the corresponding
tensor structure, which will be called the {\it tensorial dimension} $\dt$.
This is defined as the highest number of free-index momenta of 
the monomials contained in the tensor
(as usual, free indices are contracted with polarizations $h_i$ or $v$).  
For instance, $\dt [ (k_1.h_2.k_2)(k_2.h_1.k_2) + k^2 q^2 (h_1.h_2)]= 4$
and $\dt [ k^2 q^2 (h_1.h_2)] = 0$. 
Therefore, the $f_0,f_2$ amplitudes in (\ref{2p-tt})
are both associated to $\dt =4$ tensors, while $f_\Lambda$ 
multiplies a $\dt =0$ expression. 

A sufficient condition for an amplitude to be scheme independent
is that its tensor has dimension $\dt > \D$:
clearly, tensors with $\dt>\D$ cannot mix with counterterms,
while $\dt \le \D$ tensors could mix with them -- a detailed analysis is 
then necessary.
For the $n$-point functions of the stress tensor ($\D=d$), 
the scheme-independence condition is $\dt > 4$ in four dimensions; all
amplitudes in $\bra TT\ket$  have associated tensors with $\dt\le 4$ and 
are actually scheme-dependent.
On the contrary, $\bra TTT\ket$ can have $\dt =6$ tensors and corresponding
finite, scheme-independent amplitudes;
finding the finite amplitudes related to the anomaly coefficients $a$ and $c$
is the main motivation for studying the three-point function.

 
\section{The stress-tensor three-point function}

\subsection{Derivation of the general form} 
 
In this Section, we derive the general form of the  
stress-tensor three-point function at the symmetric point
$k^2\equiv k^2_1=k^2_2$:  
\beq\label{3p} 
 \bra \left(h_3. T(k_3) \right)  
\left(h_2. T(k_2) \right) \left(h_1. T(k_1) \right)\ket \ , 
 \qquad q^\mu\equiv -k_3^\mu=k_1^\mu + k_2^\mu\ ,  
\qquad k^2_1=k^2_2\equiv k^2\ . 
\eeq 
Although the point $q^2=k^2$ is the most symmetric, it will be not 
consider here because it lays in the unphysical region 
of complex momenta, while the choice $q^2 \neq k^2$ allows
for real (Minkowskian) momenta both at and off criticality. 
 
The diffeomorphism Ward identity constraints the longitudinal component
of the stress tensors: for $T(k_3)$,  it reads 
(see Appendix \ref{app-ward} for the derivation):
\bea\label{3p-diff} 
 0 & = & 
2 \bra (k_3.T(k_3).v)(T(k_2).h_2)(T(k_1).h_1) \ket  \nl  
 & + & 2 \bra (k_3.h_2.T(-k_1).v)(T(k_1).h_1) \ket  
- (k_2.v) \bra (T(-k_1).h_2)(T(k_1).h_1) \ket  \nl  
& + & 2 \bra (k_3.h_1.T(-k_2).v)(T(k_2).h_2) \ket  
- (k_1.v) \bra (T(-k_2).h_1)(T(k_2).h_2) \ket \ .  
\eea 
Note that this inhomogeneous equation relates the three-point function
to the two-point correlator already found in Section 2.3.


\begin{table}
\begin{center}
\begin{tabular}{|c||c|c|c|c|}
\hline
{\bf N}
&
{\bf  N+1}
&
{\bf N+2}
&
{\bf N+3}
&
{\bf N+4}
\\
\hline\hline
$
0
$
&
$
(1|1|1)\,(1|2|1)\,(1|3|1)
$
&
$
(1|3|1)\,(2|1|2)\,(2|2|2)
$
&
$
(1|1|1)\,(1|2|1)\,(1|3|2)
$
&
$
(1|1|1)\,(1|3|1)\,(2|2|2)
$
\\
\hline
$
4
$
&
$
(1|1|1)\,(1|3|2)\,(2|2|2)
$
&
$
(1|1|1)\,(1|3|1)\,(1|2|2)
$
&
$
(1|3|1)\,(1|1|2)\,(2|2|2)
$
&
$
(1|1|1)\,(1|2|2)\,(1|3|2)
$
\\
\hline
$
8
$
&
$
(1|2|1)\,(1|3|1)\,(2|1|2)
$
&
$
(1|2|1)\,(1|3|2)\,(2|1|2)
$
&
$
(1|3|1)\,(1|2|2)\,(2|1|2)
$
&
$
(1|2|1)\,(1|3|1)\,(1|1|2)
$
\\
\hline
$
12
$
&
$
(1|2|1)\,(1|1|2)\,(1|3|2)
$
&
$
(1|3|1)\,(1|1|2)\,(1|2|2)
$
&
$
(1|1|2)\,(1|2|2)\,(1|3|2)
$
&
$
(1|1|1)\,(1|32|1)
$
\\
\hline
$
16
$
&
$
(1|31|1)\,(2|2|2)
$
&
$
(1|1|1)\,(1|32|2)
$
&
$
(1|1|1)\,(2|32|1)
$
&
$
(1|32|1)\,(2|1|2)
$
\\
\hline
$
20
$
&
$
(1|2|1)\,(1|31|1)
$
&
$
(1|2|1)\,(2|31|1)
$
&
$
(1|2|1)\,(1|31|2)
$
&
$
(1|32|1)\,(1|1|2)
$
\\
\hline
$
24
$
&
$
(1|31|1)\,(1|2|2)
$
&
$
(1|1|2)\,(1|32|2)
$
&
$
(1|2|2)\,(1|31|2)
$
&
$
(1|3|1)\,(1|21|1)
$
\\
\hline
$
28
$
&
$
(1|21|1)\,(2|3|2)
$
&
$
(1|3|1)\,(1|21|2)
$
&
$
(1|3|1)\,(2|21|1)
$
&
$
(1|21|1)\,(1|3|2)
$
\\
\hline
$
32
$
&
$
(1|3|2)\,(1|21|2)
$
&
$
(1|3|2)\,(2|21|1)
$
&
$
(3)\,(1|1|1)\,(1|2|1)
$
&
$
(3)\,(1|1|1)\,(2|2|2)
$
\\
\hline
$
36
$
&
$
(3)\,(1|1|1)\,(1|2|2)
$
&
$
(2)\,(1|1|1)\,(1|3|1)
$
&
$
(1)\,(1|3|1)\,(2|2|2)
$
&
$
(2)\,(1|1|1)\,(1|3|2)
$
\\
\hline
$
40
$
&
$
(3)\,(1|2|1)\,(2|1|2)
$
&
$
(3)\,(1|2|1)\,(1|1|2)
$
&
$
(2)\,(1|3|1)\,(2|1|2)
$
&
$
(1)\,(1|2|1)\,(1|3|1)
$
\\
\hline
$
44
$
&
$
(1)\,(1|2|1)\,(1|3|2)
$
&
$
(3)\,(1|1|2)\,(1|2|2)
$
&
$
(2)\,(1|3|1)\,(1|1|2)
$
&
$
(1)\,(1|3|1)\,(1|2|2)
$
\\
\hline
$
48
$
&
$
(2)\,(1|1|2)\,(1|3|2)
$
&
$
(1|312|1)
$
&
$
(1|321|1)
$
&
$
(1|231|1)
$
\\
\hline
$
52
$
&
$
(1|312|2)
$
&
$
(1|321|2)
$
&
$
(1|231|2)
$
&
$
(2|231|1)
$
\\
\hline
$
56
$
&
$
(32)\,(1|1|1)
$
&
$
(31)\,(1|2|1)
$
&
$
(32)\,(1|1|2)
$
&
$
(1)\,(1|32|1)
$
\\
\hline
$
60
$
&
$
(2)\,(1|31|1)
$
&
$
(1)\,(1|32|2)
$
&
$
(2)\,(1|31|2)
$
&
$
(21)\,(1|3|1)
$
\\
\hline
$
64
$
&
$
(21)\,(1|3|2)
$
&
$
(3)\,(1|21|1)
$
&
$
(3)\,(1|21|2)
$
&
$
(3)\,(2|21|1)
$
\\
\hline
$
68
$
&
$
(2)\,(3)\,(1|1|1)
$
&
$
(1)\,(3)\,(1|2|1)
$
&
$
(2)\,(3)\,(1|1|2)
$
&
$
(1)\,(2)\,(1|3|1)
$
\\
\hline
$
72
$
&
$
(1)\,(2)\,(1|3|2)
$
&
$
(321)
$
&
$
(2)\,(31)
$
&
$
(3)\,(21)
$
\\
\hline
$
76
$
&
$
(1)\,(2)\,(3)
$
&

&

&

\\
\hline
\end{tabular}


\caption{The basis of $(1\leftrightarrow 2)$ symmetric six-index polynomials
of $k_1^\mu,k_2^\nu$ and $\eta_{\a\b}$:
${\cal P}_i={\cal P}_i(k_1,h_1,k_2,h_2,h_3)$, $i=1,\dots,77$.
 We use the short-hand notations:
 $(i|abc|j)\equiv k_i.h_a.h_b.h_c.k_j$,...,
 $(i|j)\equiv k_i.k_j$
 $(abc)\equiv {\rm tr} (h_a.h_b.h_c)$,...,
 $(a)\equiv {\rm tr} (h_a)$
 and we omit the $(1\leftrightarrow 2)$
 exchanged term that must be added to all
 non $(1\leftrightarrow 2)$ symmetric polynomials of the table.
The tensorial dimension of the polynomial ${\cal P}_i$ is:
$\dt=6$ for $i=1,\dots,15$, $\dt=4$ for $i=16,\dots,49$; 
$\dt=2$ for $i=50,\dots,73$ and $\dt=0$ for $i=74,\dots,77$. }
\label{t_base}
\end{center}
\end{table}

The general form of $\TTT$ off criticality can be expanded over the
basis of the Lorentz tensors which are built out of 
$k_1^\a,k_2^\b,\eta^{\mu\nu}$:
there are $137$ elementary polynomials
${\cal P}_i={\cal P}_i(k_1^\a,k_2^\b,\eta^{\mu\nu})$  
of maximal tensorial dimension $\dt=6$,
which are classified in Appendix \ref{app-tens}.
After imposing the Bose 
symmetry $(h_1,k_1) \leftrightarrow (h_2,k_2)$ at the symmetric point
$k^2_1=k^2_2$, the number of independent polynomials reduces to $77$
(see Table \ref{t_base}).
This general expansion is plugged into the Ward identity 
(\ref{3p-diff}) and the corresponding one for $(k_1 \leftrightarrow k_3)$, 
which yield linear systems for the scalar amplitudes
(the algebra is done with the help of Mathematica \cite{wolf} routines). 
In order to avoid the problem of possible degeneracies of 
the tensor basis due to relations proper to four-dimensions,
we solve the system in arbitrary large dimensions,
and postpone the discussion of the limit to four dimension. 
The solution of the Ward identity (\ref{3p-diff}) and its 
($1 \leftrightarrow 3$) analogue
contains $17$ transverse tensors, 
solutions of the homogeneous equations, plus $5$ inhomogeneous 
(i.e. non transverse) tensors multiplying the two- and one-point amplitudes
$\left\{ f_0\left(q^2 \right), f_0\left(k^2 \right),
f_2\left(q^2 \right), f_2\left(k^2 \right), f_\Lambda \right\}$
already found in Section 2.3:
\bea\label{3p-cons} 
\bh\bh\bra (h_1.T)(h_2.T)(h_3.T)\ket &=& 
     \sum_{i=1}^{17} A_i(q^2,k^2)\ {\cal T}_i  \nl
     &+& f_0\left(q^2 \right)\ {\cal T}_{0q} + 
         f_0\left(k^2 \right)\ {\cal T}_{0k} 
      +  f_2\left(q^2 \right)\ {\cal T}_{2q} + 
         f_2\left(k^2 \right)\ {\cal T}_{2k} \nl
     &+& f_\Lambda\ {\cal T}_\Lambda \ . 
\eea 
Note that the tensor structures are contracted with 
the polarizations $(h_1,h_2,h_3)$, i.e. 
${\cal T}_\omega ={\cal T}_\omega (k_1,h_1,k_2,h_2,h_3)$,
for $\o=1,\dots,17,0q,0k,2q,2k,\Lambda$.
 
At criticality  we must impose the Ward 
identities for the Weyl symmetry, which compare $\TTT$ traced
over one of the stress tensors with the second variation of 
the anomaly (\ref{1p-weyl}). There are two independent equations by
tracing $T(k_3)$ or $T(k_1)$; in the first case, we find (see Appendix B):
\bea\label{3p-weyl} 
&&\bh\bh \bra \T(k_3)(T(k_1).h_1)(T(k_2).h_2) \ket \nl 
&+& \left[ \bra (T(-k_2).h_1)(T(k_2).h_2) \ket + 
                     \bra (T(k_1).h_1)(T(-k_1).h_2) \ket \right] \nl 
&= & \A (k_1,h_1;k_2,h_2) - \half \left[ 
(h_1) \A (k_2,h_2) + (h_2) \A (k_1,h_1) \right] \ .
\eea 
In this Equation, $\A (k_1,h_1)$ and $ \A (k_1,h_1;k_2,h_2)$ are, 
respectively, the first and second variation of the anomaly (\ref{1p-weyl})  
w.r.t. to the metric, $\left(h_i\ .\ \d /\d g(k_i) \right)$,  
evaluated in flat space. 
Their explicit expressions are:
\beq 
\A (k_1,h_1)/\l = a' D^2 {\cal R}(k_1,h_1) = a' 
k_1^2\left( k_1^2 (h_1) -(k_1.h_1.k_1) \right) \ ; 
\eeq 
and 
\bea 
\A (k_1,h_1;k_2,h_2)/\l &=& \left( -3 c + a \right)  
            \ \left( R_{\mu\nu\a\b} \right)^2(k_1,h_1;k_2,h_2) \nl 
& + & \left(6  c - 4 a \right) 
          \ \left({\cal R}_{\a\b} \right)^2(k_1,h_1;k_2,h_2) \nl 
& + & \left(- c + a + r \right)\ {\cal R}^2(k_1,h_1,k_2,h_2) \nl 
& + & a'\ D^2 {\cal R}(k_1,h_1;h_2,k_2)\ , 
\eea 
with 
\beq
\begin{array}{lll} 
\left( R_{\mu\nu\a\b} \right)^2(k_1,h_1;k_2,h_2) 
&=& 
2\left[ (k_1.k_2)^2 (h_1.h_2) +(k_1.h_2.k_1)(k_2.h_1.k_2) \right.
\\
&&\left. - 2 (k_1.k_2) (k_1.h_2.h_1.k_2) \right] ,
\\ 
\left({\cal R}_{\a\b} \right)^2(k_1,h_1;k_2,h_2)
&=& 
{1\over 2} (k_1.k_2)^2 (h_1)(h_2) + {1\over 2} k_1^2 k_2^2 (h_1.h_2) 
\\
&&+  
(k_1.k_2)(k_1.h_1.h_2.k_2) + (k_2.h_2.k_1)(k_1.h_1.k_2)
\\
&&+  
\left[ {1\over 2} k_2^2 (h_1)(k_1.h_2.k_1) - 
(k_1.k_2) (h_1) (k_2.h_2.k_1) \right. 
\\
&& \left. - k_1^2 (k_2.h_2.h_1.k_2) + 
\left\{ 1 \leftrightarrow 2 \right\} \right] ,
\\
{\cal R}^2(k_1,h_1;k_2,h_2) 
&=& 2\left[ \left( k_1^2 (h_1) - (k_1.h_1.k_1) \right) 
\left( \left\{ 1 \leftrightarrow 2 \right\} \right)\right] ,
\\
D^2 {\cal R}(k_1,h_1;k_2,h_2)
&=& 
\left[ \left( k_1^2 (h_1) - (k_1.h_1.k_1) \right) \right. 
\\ 
&&\times \left. \left( 
(k_1.h_2.k_1) + (k_2.h_2.k_1) -{1\over 2} (k_1.k_2) (h_2) \right)
+ \left( \left\{ 1 \leftrightarrow 2 \right\} \right) \right] 
\\
&&+ (k_1+k_2)^2 \left[- (h_1.h_2) 
\left( k_1^2 + k_2^2 +{5\over 2} (k_1.k_2) \right)
- {1\over 2} (k_1.k_2) (h_1)(h_2) \right.
\\
&&+ (k_2.h_1.h_2.k_1) + (h_1) \left( (k_1.h_2.k_1) + (k_1.h_2.k_2) \right)
\\
&&+ \left. (h_2) \left( (k_2.h_1.k_2) + (k_2.h_1.k_1) \right) \right] .
\end{array}
\label{var-a}\eeq
The consistency among the two Weyl Ward identities
implies the vanishing of the coefficient $r$ in the trace anomaly 
(\ref{theta-def}) at criticality; this is
the Wess-Zumino condition, which
requires the commutativity of Weyl variations \cite{chi-book}. 
Furthermore, the consistencies between Weyl and Diff identities imply 
$f^c_0\left(q^2 \right)=f^c_0\left(k^2 \right)= \l \ a'/3$  
in agreement with the earlier result (\ref{f0}).   
 
Next, the two Weyl Ward identities are imposed on the form of 
$\TTT$ satisfying the diffeomorphism identities, Eq.(\ref{3p-cons});
a solution is found in four dimensions only, and
the number of independent tensor structures reduces from $22=(17$
transverse $+5$ non-transverse) to 
$13=(8$ transverse-traceless $+2$ transverse-traceful 
$+3$ non-transverse-traceful).
The final result for $\bra TTT \ket$ off-criticality is thus given by
(\ref{3p-cons}) up to a convenient change of basis, which
makes the critical limit simpler and also fulfills other requirements
to be explained later:
\bea
\label{3p-end} 
\bh\bh\bra (h_1.T)(h_2.T)(h_3.T)\ket 
&=& 
\sum_{i=1}^8 A_i(q^2,k^2)\ {\cal T}_i
+ A_E(q^2,k^2)\ {\cal T}_E + A_W (q^2,k^2)\ {\cal T}_W \nl
&+& 
\sum_{i=11}^{17} A_i(q^2,k^2)\ {\cal T}_i \nl
&+&
\sum_{\pm} \left[ f_{0\pm}\left(q^2,k^2\right)\ {\cal T}_{0\pm} +
        f_{2\pm}\left(q^2,k^2\right)\ {\cal T}_{2\pm} \right]
\ +\ f_\Lambda\ {\cal T}_\Lambda \ .
\eea 
In this Equation, the $22$ tensors ${\cal T}_\omega$ 
only depend on the external momenta and do not contain any parameter
of the theory, thus they are invariant of the renormalization-group
flow and provide a convenient basis both at and off criticality.
The scalar amplitudes $A_i$ carry all the dependence on the flow:
at criticality, some of them are constrained by the Weyl identities,
in particular $9=22-13$ of them vanish $(7$ transverse-traceful $+2 $ 
non-transverse-traceful).
This is made manifest in the basis in Eq. (\ref{3p-end}), which 
is an extension of the basis at criticality.


\begin{table}
\begin{center}
 \newcommand\Tc[1]{{\cal T}^c_{#1}}
 \newcommand\Tm[1]{{\cal T}_{#1}}
 \newcommand\Tb[1]{t_{#1}}
 \begin{tabular}{|c||c|c|c|c|c|c|}
\hline
&
$\bot k_{1,2}$
&
$\bot k_3$
&
${\rm tr}_{1,2}=0$
&
${\rm tr}_{3}=0$
&
$\delta_T$
&
$\Delta$
\\
\hline\hline
$
\Tm{1}
$
&
y
&
y
&
y
&
y
&
$
6
$
&
$
> 6
$
\\
\hline
$
\Tm{2}
$
&
y
&
y
&
y
&
y
&
$
6
$
&
$
> 6
$
\\
\hline
$
\Tm{3}
$
&
y
&
y
&
y
&
y
&
$
6
$
&
$
> 6
$
\\
\hline
$
\Tm{4}
$
&
y
&
y
&
y
&
y
&
$
6
$
&
$
> 6
$
\\
\hline
$
\Tm{5}
$
&
y
&
y
&
y
&
y
&
$
6
$
&
$
> 6
$
\\
\hline
$
\Tm{6}
$
&
y
&
y
&
y
&
y
&
$
6
$
&
$
> 6
$
\\
\hline
$
\Tm{7}
$
&
y
&
y
&
y
&
y
&
$
6
$
&
$
> 6
$
\\
\hline
$
\Tm{8}
$
&
y
&
y
&
y
&
y
&
$
4
$
&
$
4
$
\\
\hline\hline
$
\Tm{E}
$
&
y
&
y
&
n
&
n
&
$
6
$
&
$
6
$
\\
\hline
$
\Tm{W}
$
&
y
&
y
&
n
&
n
&
$
6
$
&
$
8
$
\\
\hline\hline
$
\Tm{11}
$
&
y
&
y
&
n
&
n
&
$
4
$
&
$
6
$
\\
\hline
$
\Tm{12}
$
&
y
&
y
&
n
&
n
&
$
4
$
&
$
6
$
\\
\hline
$
\Tm{13}
$
&
y
&
y
&
n
&
n
&
$
6
$
&
$
6
$
\\
\hline
$
\Tm{14}
$
&
y
&
y
&
n
&
n
&
$
6
$
&
$
6
$
\\
\hline
$
\Tm{15}
$
&
y
&
y
&
n
&
n
&
$
6
$
&
$
6
$
\\
\hline
$
\Tm{16}
$
&
y
&
y
&
n
&
n
&
$
6
$
&
$
6
$
\\
\hline
$
\Tm{17}
$
&
y
&
y
&
n
&
n
&
$
6
$
&
$
> 6
$
\\
\hline\hline
$
{\cal T}_{0+}
$
&
n
&
n
&
n
&
n
&
$
4
$
&
$
4
$
\\
\hline
$
{\cal T}_{0-}
$
&
n
&
n
&
n
&
n
&
$
6
$
&
$
>6
$
\\
\hline
$
{\cal T}_{2+}
$
&
n
&
n
&
n
&
n
&
$
4
$
&
$
4
$
\\
\hline
$
{\cal T}_{2-}
$
&
n
&
n
&
n
&
n
&
$
6
$
&
$
6
$
\\
\hline
$
{\cal T}_\Lambda
$
&
n
&
n
&
n
&
n
&
$
0
$
&
$
4
$
\\
\hline\hline
\end{tabular}

\caption{Properties of the tensors in Eq.(\ref{3p-end}); ($\bot$):
transversality w.r.t. $k_1(k_2)$ and $k_3$ (yes/no);  (${\rm tr}=0$):
tracelessness w.r.t. $T_1(T_2)$ and $T_3$; $\d_T$: tensorial dimension;
$\Delta$: scale dimension of the tensor in polynomial form.}
\label{t_prop}
\end{center}
\end{table}

More specifically, the $22$ tensors have
the following properties (see also Table \ref{t_prop}):
\begin{itemize}
\item
The tensors ${\cal T}_i$, $i=1,\dots,8$ are transverse and traceless
w.r.t. all the three stress-tensors and thus the corresponding
amplitudes are unconstrained at criticality.
\item
The tensors ${\cal T}_E\equiv {\cal T}_9$ and 
${\cal T}_W\equiv {\cal T}_{10}$ are transverse and traceful;
their traces match the second variation (\ref{var-a}) of the 
Euler and Weyl-square terms in the anomaly(\ref{theta-def}), respectively; 
the corresponding amplitudes have the critical limits:
\bea
A_E(q^2,k^2)& \to& A^c_E(q^2,k^2) = \l\ \frac{a}{q^2}\ , \nl 
A_W(q^2,k^2)& \to& A^c_W(q^2,k^2)\ =\l\ \frac{c}{q^2} \ ,\qquad
{\rm (critical\ \ points)}.
\label{EW-lim}
\eea
\item
The seven tensors ${\cal T}_i$, $i=11,\dots,17$ are transverse and
traceful, but their traces do not match any of the terms in the anomaly
(\ref{theta-def}). Therefore, their amplitudes should vanish at criticality:
\beq
A_i(q^2,k^2)\ \to\ 0\ , \quad i=11,\dots,17, \qquad
\qquad {\rm (critical\ \ points)}.
\label{ai-lim}
\eeq 
These amplitudes are nevertheless different from zero off criticality;
another parameterization of $\bra \Theta \ket$ off criticality has been given 
in Ref.\cite{jo}, but cannot be immediately compared with our result.
\item
The five tensors ${\cal T}_{i\pm}$, $i=0,2$, and ${\cal T}_\Lambda$
are neither transverse nor traceless; the first four are the even and
odd combinations of the tensors multiplying the two-point
amplitudes $f_i\left(q^2\right)$ and $f_i\left(k^2\right)$ 
(see Eq.(\ref{3p-cons})):
\bea
{\cal T}_{i\pm}&\equiv &
\left( {\cal T}_{iq} \pm {\cal T}_{ik} \right)\ ,\label{pm-tens}\\
f_{i\pm} &\equiv & \frac{1}{2}
\left( f_i\left(q^2\right) \pm f_i\left(k^2\right) \right)\ ,
\qquad i=0,2\ .
\label{pm-amp}
\eea
The fifth tensor accounts for the one-point amplitude $f_\Lambda$
(\ref{1p-t}). The critical limits of these amplitudes are:
\bea
f_{0+}\left(q^2,k^2 \right) & \to & \l\ \frac{a'}{3}\ , 
\label{f0p}\\
f_{0-}\left(q^2,k^2 \right) & \to & 0 \ , 
\label{f0m}\\
f_{2\pm}\left(q^2,k^2 \right) & \to & - \l \frac{c}{2}
\left(  \log\left(q^2\over \mu^2 \right) \pm
        \log\left(k^2\over \mu^2 \right)
       \right)\ , \label{f2pm} \\
f_\Lambda & \to & 0\ ,\qquad\qquad\qquad\qquad {\rm (critical\ \ points).} 
\label{flac}
\eea
Actually, the solution of the Weyl identities (\ref{3p-weyl}) 
yields the results (\ref{f0p}, \ref{f0m}), which agree with the
earlier result (\ref{f0});
on the other hand, the spin-two amplitudes $f_{2\pm}$ are not
constrained by the Weyl identities and their critical values are
taken from (\ref{f2}).
\end{itemize}


\subsection{Further properties and renormalization} 

{\bf Null tensor.}\\
The basis of elementary tensors $\{ {\cal P}_i\}$ (Table \ref{t_base}) 
is degenerate in four dimensions by the same mechanism found for
$\bra AVV\ket$ (Eq.(\ref{rose-ide})) and the two-dimensional
$\bra TT\ket$: 
there exists one ``null tensor'' which do not manifestly vanish in 
the tensor notation used here, but actually has all
components equal to zero in four dimensions.
Such six-index Bose-symmetric null vector 
was found by inspection and, moreover, 
no null vectors were found with five and four indices
within the tensors bases used for the Diff and Weyl Ward identities,
respectively.
Therefore, the corresponding linear systems for the coefficients
have been correctly extracted from faithful bases. 
The null tensor clearly satisfies all Ward identities and can be eliminated
{\it a-posteriori} from the final result (\ref{3p-end}):
it is the traceless-transverse tensor ${\cal T}_8$
(see Table \ref{t_prop}), which will be discard hereafter
(its expression is given in Appendix D).

\bigskip

\noindent
{\bf Choice of basis for the tensors.}\\
Each of the tensors ${\cal T}_\omega$ can be considered as a 
vector ($u$) in the 
$77$-dimensional ${\cal P}_i$-space (Table \ref{t_base}) and the
Ward identities are inhomogeneous linear equations
of the form $A\cdot u=w$, whose solutions possess characteristic
ambiguities:
(I) the homogeneous solutions $u_0$, $\ A\cdot u_0=0$, are determined 
up to linear combinations and must be normalized;
(II) the inhomogeneous solutions are defined up to
the addition of homogeneous solutions: 
$u=A^{-1}\cdot w \ \to \ A^{-1}\cdot w +u_0$.
Using this freedom, we can choose a basis of tensors
${\cal T}_\omega$ which is more suitable for the physical interpretation.
Generically, the tensors have the form:
\beq
{\cal T}_\omega = \sum_{i=1}^{77} \ Q_i^\o(q^2,k^2) 
\ {\cal P}_i \left(k_1,h_1,k_2,h_2,h_3\right)\ ,
\label{polyform}
\eeq
where the $Q_i^\o$ are rational functions of the Lorentz invariants
with various scale dimensions (the ${\cal P}_i$ have dimension equal
to their tensorial dimension, $\Delta=\dt =0,2,4,6$, listed in Table
\ref{t_base}). Thus, the two ambiguities (I) and (II)
can be used to find convenient forms for the $Q_i^\o$.

Let us first consider the choice in which the $Q_i^\o$ are polynomial:
this is clearly possible for the homogeneous solutions of the 
Diff Ward identity (${\cal T}_i$, $i=1,\dots,17$), 
by factoring out the common denominators into the corresponding amplitudes. 
As a matter of fact, four of the inhomogeneous solutions
(last five entries in Table \ref{t_prop}) can also be put in 
polynomial form using type-(II) redefinitions and without 
changing their amplitudes, which are inputs of the Ward identities. 
The scale dimensions of the ${\cal T}_\omega$'s in polynomial form are listed
in the last column of Table \ref{t_prop}; we find the following cases:
\begin{itemize}
\item
The tensors with $\Delta =4$ and $\dt \le 4$  are naturally polynomial
because they have the same dimension as $\bra TTT\ket$;
the corresponding amplitudes are dimensionless functions (or constants),
as in the case of the two-point function (Section 2.3).
\item
The polynomial form also makes sense for the tensors with 
$\Delta=6$ and $\dt=4,6$; their amplitudes have the natural dimension $-2$
and may display a simple pole in the massless theory.
\item
The polynomial form is not convenient for
tensors with $\dt=6$ and $\Delta >6$;
the corresponding amplitudes would have large negative dimensions 
$(4-\Delta=-4,-6)$ which suggest the possibility of unphysical 
higher-order poles to be cancelled by corresponding zeroes in the tensors. 
By inspection, we have found another non-polynomial form
for these tensors which is characterized by scale dimension $6$ 
and rational $Q_i^\o$ 
with simple poles $\sim 1/(q^2 - 2 k^2 )$, $1/(q^2 + 2 k^2 )$
(higher-order poles and $1/k^2$ poles are absent in this basis).
The corresponding amplitudes have again dimension $-2$.
\end{itemize}
In conclusion, the general $\bra TTT\ket$ can be put into a
form which does not explicit contain higher-order poles 
and possess a smooth limit\footnote{ 
Note that all tensors become polynomial at $k^2=0$; the solution of the
Ward identities at $k^2=0$  will be further discussed in Section 4.}
$k^2\to 0$ with $q^2\neq 0$.
This form is convenient in the kinematical region $q^2 > 2k^2 \ge 0$;
other choices of basis may be relevant for different corners
of the phase space, say $q^2 \to 0$ with $k^2\neq 0$. 

\bigskip

\noindent
{\bf Renormalization and scheme-dependence of the amplitudes.}\\
As discussed at the end of Section 2.3, the form of $\TTT$ in 
(\ref{3p-end}) can be changed by adding counterterms, which are
polynomials of the momenta of maximal dimension four.
The counterterms cannot be arbitrary, because they would change the
definition of the stress tensor, e.g. by transforming a trace anomaly into
a gravitational anomaly; therefore, the counterterms should be already present
in the solution of the Ward identities just found.
We can identify them as the tensors with $\Delta=4$ and $\dt \le 4$,
which actually are polynomials; the corresponding
dimensionless amplitudes are scheme-dependent by the shifts
$A_\o \to A_\o + const.$.
Such tensors are:
\begin{itemize}
\item
The null tensor ${\cal T}_8$ which vanishes identically.
\item
The tensor ${\cal T}_{0+}$ which matches the
scheme-dependent $D^2 {\cal R}$ term in the trace anomaly; 
this verifies the ambiguity
$a' \to a' + const.$ at the level of the three-point function.
\item
The tensor ${\cal T}_{2+}$, which accounts
for the $\mu$-dependence of the logarithms in the amplitude
$f_{2+}$ (Eq.(\ref{f2pm})); such dependence instead cancels out in the
odd combination $f_{2-}$, which multiplies a $\Delta=\dt=6$ tensor.
\item
The tensor ${\cal T}_\Lambda$ coming from the one-point function off
criticality.
\end{itemize}
These results are in agreement with earlier analyses based on the
lower-point functions and the effective action \cite{duff}\cite{cfl}
\cite{ds}\cite{os}, and provide a check of our derivation.

Note that two further tensors, ${\cal T}_{11}$ and
${\cal T}_{12}$, have dimensions $\dt =4 $ and $\Delta=6$,
so that they cannot mix with counterterms. 
Here we have checked that their dimension cannot be further lowered by
changing the basis: the mixing with other tensors is not actually 
possible owing to their higher $\dt$ value.

In conclusion, the $\dt=6$ tensors in Table \ref{t_prop} 
(and the previous ${\cal T}_{11}$ and ${\cal T}_{12}$) identify
amplitudes $A_i(q^2, k^2)$ which are finite and scheme-independent 
with scale dimension $(-2)$.
In particular, we found the desired result of 
matching the (universal) anomaly coefficients $a,c$
with two of these amplitudes, namely $A_E$ and $A_W$ in (\ref{EW-lim}).

\bigskip

\noindent
{\bf Limit of the solution for $k^2\to 0$}\\
As already said, the limit is smooth in our choice of ${\cal T}_\omega$ 
basis; however, one tensor becomes a linear combination of the others
and should be removed.
Actually, for $k^2=0$, the two tensor structures of the two-point function 
$\bra \left(h_1.T(k)\right) \left(h_2.T(-k)\right)\ket$
are degenerate (see Eq.(\ref{2p-tt})), so that the Diff
Ward identity (\ref{3p-diff}) looses one inhomogeneous solution.
Moreover, the amplitude $f_2\left(k^2\right)$ becomes singular
at criticality (see Eq.(\ref{f2})); thus, the corresponding term 
$f_2\left(k^2\right)\left[ {\cal T}_{2+}-{\cal T}_{2-} \right]$
is subtracted from $\bra TTT\ket$:
\bea
&& \bra (h_1.T)(h_2.T)(h_3.T)\ket_{{\rm reg},\ k^2=0} \nl
&& \equiv 
\lim_{k^2\to 0} \left\{ \bra (h_1.T)(h_2.T)(h_3.T)\ket 
- \frac{f_2\left(k^2\right)}{2}\left[ {\cal T}_{2+}-{\cal T}_{2-} \right]
\right\} \ .
\label{k0lim}
\eea
The $k^2=0$ solution thus contain $21=(17$ transverse $+4$
non-transverse) independent tensors. The analysis of the Ward identities
discussed before has been independently repeated at the point $k^2=0$ 
and matched with the result of the limit from $k^2\neq 0$. 
A summary of the properties of the solutions for both $k^2=0$ and 
$k^2\neq 0$ is given in Table \ref{t_sol}; note that
some of the tensors have lower $\dt$ values at $k^2=0$,
but the number and type of homogeneous solutions remains the same.

\bigskip


\begin{table}
\begin{center}
\begin{tabular}{|c|c|c|c|c|c|c|c|c|c|c|c|c|}
\hline
  & 
\multicolumn{5}{|c|}{ transverse} & 
\multicolumn{5}{c|}{non-transverse} & 
\multicolumn{2}{c|}{total} 
\\
\hline\hline
$ k^2 \neq 0 $ &
\multicolumn{3}{|c|}{$ \dt =6 $} & \multicolumn{2}{c|}{$ \dt =4 $} &
\multicolumn{2}{c|}{$ \dt =6 $}  & \multicolumn{3}{c|}{$ \dt \le 4 $} &
                      $ \dt =6 $ & $ \dt \le 4 $
\\ 
\hline 
 Diff & \multicolumn{3}{|c|}{ 14 } & \multicolumn{2}{c|}{ 2 (3)} &
      \multicolumn{2}{c|}{ 2 }     & \multicolumn{3}{c|}{ 3 } & 
                                16 & 5 
\\
Weyl    &   7      &   2       &  5      &   (1)        &  2       &
  1     & 1        &  1        & 1       & 1          & 10       & 2    
\\
{\it value}  &$  (tr=0)  $&$ a, c    $&  0      &  null      & 0        &
$ f^c_{2-}$&  0     &$   f^c_{2+} $&$ a'    $&$ 0 $&          &    
\\
{\it label i} &  1,..,7 & 9,10 & 13,..,17 &  8  & 11,12      &
$2-     $&$ 0-   $& 2+         &   0+    &$ \Lambda$  &          &          
\\
\hline\hline
$ k^2 = 0 $ &
\multicolumn{3}{|c|}{$ \dt =6 $} & \multicolumn{2}{c|}{$ \dt =4 $} &
\multicolumn{2}{c|}{$ \dt =6 $}  & \multicolumn{3}{c|}{$ \dt \le 4 $} &
                      $ \dt =6 $ & $ \dt \le 4 $
\\ 
\hline 
 Diff & \multicolumn{3}{|c|}{ 12 } & \multicolumn{2}{c|}{ 4 (5)} &
      \multicolumn{2}{c|}{ 2 }     & \multicolumn{3}{c|}{ 2 } & 14 & 6 
\\
Weyl    &   7      &   2       &  3      &   (1)        &  4       &
1       &  1        &\multicolumn{2}{c|}{ 1} & $ 1 $       & 10  & 1    
\\
{\it value} &$  (tr=0)  $&$ a, c    $&  0      &  null      & 0        &
$ f^c_2(q^2)$ & 0   &\multicolumn{2}{c|}{ $a'$} &$ 0 $&       &    
\\
{\it label i} &  1,..,7 & 9,10 & 15,..,17 &  8  & 11,..,14   &
$2$    &  $ 0-$   & \multicolumn{2}{c|}{ $0+$}  & $ \Lambda$  &    &     
\\
\hline\hline
\end{tabular}

\caption{Summary of the properties of the solutions to the Ward identities
for $k^2\neq 0$ and $k^2 = 0$. Using the notations of Table \ref{t_prop},
we report the number of solutions, split by the $\dt$ value,
both off criticality ({\it Diff} line) and
at criticality ({\it Weyl}); we also indicate
the critical limit of the amplitudes ({\it value}) and their labels.}
\label{t_sol}
\end{center}
\end{table}


\subsection{Critical points: conformal isometries of flat space}

In the previous Section, we obtained the general form (\ref{3p-end}) of 
$\TTT$, which contains 9 scheme-independent amplitudes at criticality,
namely $A_i, i=1,\dots,7$, $A_E$ and $(A_W, f_{2-})$.
Previous approaches in coordinate space 
have instead found three independent non-contact tensor structures
\cite{eo}\cite{ol}: two parameterized by the $a$ and $c$ anomalies 
and one traceless transverse\footnote{
The three independent structures were originally found in free field theory
by the first of Refs.\cite{eo}, and
were confirmed by the subsequent analyses of the conformal symmetry.
}.
The disagreement is due to the fact that,  
in our derivation, we did not impose 
the symmetry under the regular (i.e. global) conformal transformations of
flat space, belonging to the group $SO(d+1,1)$ \cite{pdf}\cite{cc};
as is well known, these symmetries almost completely determine
the functional form of the two- and three-point functions in coordinate 
space. Moreover, we did not considered the functional relations 
among amplitudes due to the exchange symmetry of $T(k_1)$ 
with $T(k_3)$.

On the other hand, the disadvantage of the coordinate-space
approach is that it yields the expressions of correlators 
for non-coincident points, which are rather difficult to
renormalize \cite{eo}. Furthermore, the coordinate approach
cannot be easily extended away from criticality
because the Diff Ward identities become involved.

In the following, we rederive the Ward identities
for the $SO(d+1,1)$ conformal symmetry in flat momentum space, without
neglecting the contact terms and consistently following the
conventions set in this paper.
As is well known \cite{cc}, a $SO(d+1,1)$ conformal transformation
is a combinations of a reparametrization and a Weyl scaling
which map the Minkowski metric into itself: the precise
notion is that of {\it conformal isometries} of the flat metric, 
which are associated to conformal Killing vectors \cite{bd}.
A general metric $g^{\mu\nu}(x)$ possesses a conformal isometry 
if it satisfies:
\beq
0=\left( \d_\eps + \d_\s \right) \ g^{\mu\nu}(x) \equiv
D^{(\mu} \eps^{\nu)}(x) -2 \s(x) \ g^{\mu\nu}(x) \ ,
\label{iso}\eeq
where $\eps^\mu$ is the conformal Killing vector;
the associated Weyl scaling  $\s(x)= (D\cdot \eps)/d$ is found
by taking the trace of (\ref{iso}).
In flat space, the solutions of (\ref{iso}) generate
the $SO(d+1,1)$ conformal group \cite{pdf}:
\beq
\eps^\mu(x)= b^\mu \ + \ \lambda x^\mu +\omega^\mu_\nu x^\nu +
a^\mu x^2 -2 (a\cdot x) x^\mu \ , \qquad\qquad (d>2).
\label{conf-eps}\eeq

We recall that $\d_\eps =- {\cal L}_\eps$ is (minus)
a Lie derivative and that the combination $\left( \d_\eps +\d_\s \right)$
is still covariant because the Weyl transformation is local.
The usual conformal Ward identities in flat space \cite{cc} are
nothing else than the covariance conditions for the fields.
For example, a scalar and a vector field satisfy:
\bea
0= - \left(\d_\eps +\d_\s \right) \varphi &=& 
\left( \eps \cdot \de \right) \varphi + 
{\D\over d} \left( \de \cdot \eps \right) \varphi \ , 
\nl
0= - \left(\d_\eps +\d_\s \right) A_\mu &=& 
\left( \eps \cdot \de \right) A_\mu + \left( \de_\mu \eps^\a \right) A_\a +
{\D-1\over d} \left( \de \cdot \eps \right) A_\mu \ . 
\label{conf-tr}\eea
The field $\varphi, A_\mu$ are said {\it quasi-primary}, because they transform
homogeneously under Weyl with scaling dimension $\D$.
In the case of the stress tensor,
we must pay attention to the effects of the trace anomaly,
and the naive (classical) formulas (\ref{conf-tr}) should not
be taken for granted.

We reconsider the Ward identities (\ref{ward-def}) for Diff and Weyl 
transformations of the generating functional and specialize them
for a conformal isometry (\ref{iso}) (for more details, see Appendix B).
The two identities can be combined to obtain the sum of
three terms: respectively, the stress-tensor $(n+1)$-function 
times $(\d_\eps +\d_\s ) g$, the $n$-th point function
and the $n$-th variation of the anomaly. 
Since the first term vanishes for conformal isometries, this
Ward identity closes on the $n$-point functions and yields 
a finite set of differential equations with inhomogeneous term
given by the anomaly.
These equations put additional constrains on the solutions
of the two separate Ward identities, 
which apply to general metrics without conformal isometries.

The conformal Ward identity for the two-point function is:
\bea
&-& \left[ \left(\d_\eps +\d_\s \right)_1
       +\left(\d_\eps +\d_\s \right)_2 \right]
 \bra T_{\a\a'}(x_1) T_{\b\b'}(x_2)\ket 
\nl
&=& -2\ {\d^2 \over \d g^{\a\a'}(x_1) \d g^{\b\b'}(x_2)}\ 
\int\sqrt{g} \ \s\ {\cal A}(g) \Bigg\vert_{\s=(\de\cdot\eps)/4}\ ,
\label{2p-conf}
\eea
where the metric variations are evaluated in flat space.
The conformal transformation of the stress tensor reads \cite{cc}:
\beq
-\left(\d_\eps +\d_\s \right) T_{\a\b} =
\left[ \left(\eps\cdot \de\right) +{d-2\over d}\left(\de\cdot\eps\right)
\right] T_{\a\b} + \left(\de_{(\a} \eps^\nu \right) T_{\nu\b)}\ .
\label{conf-t}\eeq

For scale variations $(\eps^\mu=b\ x^\mu)$
and in four-dimensional momentum space, Eq.(\ref{2p-conf}) becomes:
\beq
\left(p\cdot \partial_p -4 \right) 
\bra(h_1. T(p))(h_2. T(-p)) \ket = 2\ {\cal A}(p,h_1;-p,h_2)\ ,
\label{c-det}\eeq
where the r.h.s. is the second variation of the integrated 
trace anomaly (\ref{theta-def}) in the notation introduced
before (see Eq.(\ref{var-a})). This variation
only contains the Weyl term, because the Euler term is topological
invariant and $\sqrt{g} D^2{\cal R}$ is a total derivative;
therefore, the r.h.s. of (\ref{c-det}) 
is a traceless-transverse tensor.
By plugging in the explicit expressions of the anomaly (\ref{var-a})
and of $\TT$ (\ref{2p-tt}), one
verifies that the coefficient of the logarithm
in spin-two amplitude $f_2\left(p^2 \right)$ is equal to the anomaly 
number $c$, as anticipated in Eq.(\ref{f2}) \cite{duff}\cite{cfl}.

For special conformal transformations, parameterized by $a_\mu$ in 
(\ref{conf-eps}), the Ward identity (\ref{2p-conf}) becomes a second-order
differential equation in momentum space; we do not solve it
but rely on earlier studies in coordinate space (see e.g. Ref.\cite{cfl})
to conclude that it is automatically satisfied by 
the critical form of $\TT$ given by (\ref{2p-tt}) 
with $f^c_0, f^c_2$ in (\ref{f0},\ref{f2}).

In conclusion, we have found that the conformal Ward identities
for the stress-tensor $n$-point correlators
contain an inhomogeneous term given by the $n$-th variation of
the trace anomaly, and that they amount to a set of differential 
equations for the scalar amplitudes at criticality. 
At the three-point level, the equations for special conformal 
transformations, together with the functional relations for the
$(1\leftrightarrow 3)$ symmetry, presumably determine the form of the
traceless transverse amplitudes, 
$A_i^c(q^2,k^2)$, $i=1,\dots,7$, leading to an
agreement with the coordinate-space result \cite{eo}\cite{ol}.
We shall not carry out this analysis, 
because we are more interested in off-critical behavior of the 
traceful terms in $\TTT$, which describe the renormalization-group
flow of the anomaly coefficients $a$ and $c$.


\section{Trace anomaly: reduction formulae and sum rules}

 
\subsection{Reduction formulae}

We now describe the inversion of the $\TTT$ expansion
(\ref{3p-end}) which extracts the amplitudes $A_\o(q^2,k^2)$ 
from an explicit expression of the three-point function,
such as the result of a calculation.
The latter can be expanded in the canonical
basis of Bose-symmetric tensors ${\cal P}_i$, $i=1,\dots,77$
(see Table \ref{t_base}),
leading to a set of know coefficient functions $u_i(q^2,k^2)$.
We then write (see also (\ref{polyform})):
\bea
\TTT & =& \sum_{\o=1}^{22}\ A_\o \ {\cal T}_\omega = 
\sum_{i=1}^{77} \sum_{\o=1}^{22}\ A_\o(q^2,k^2) \ Q^\o_i(q^2,k^2) 
\ {\cal P}_i \left(k_1,h_1,k_2,h_2,h_3\right) \nl
&=& \sum_{i=1}^{77} \ u_i(q^2,k^2)\ {\cal P}_i \ .
\label{3p-inv}
\eea
After the elimination of the null vector,
the $21$  ${\cal T}_\o$'s and the $76$ ${\cal P}_i$'s form
two sets of linearly-independent ``vectors''; therefore,
Equation (\ref{3p-inv}) defines the linear
problem of finding the ``coefficients'' $A_\o$ by
inverting the 21x76 matrix $Q^\o_i$ of maximal rank.
The solution is unique and expresses the $21$ 
$A_\o$'s as linear combinations of the $u_i$'s, together with
$55$ linear relations among the $u_i$'s.

These constraints verify that the calculation, besides being
correct, satisfies the Diff Ward identities in our specific
formulation, which includes a complete choice of renormalization
scheme and of stress-tensor definitions.
This leads to the practical problem that a given 
calculation may not immediately fit our scheme, and thus the 
check of our Ward identities (i.e. of the 55 constraints) may 
require modifications of the calculation rules and become a 
painful process. 
For example, in Section 5 it is shown that 
a Feynman-graph calculation of $\TTT$ needs
the inclusion of $T_{\mu\nu}$ vertices with an increasing number of
legs in order to match our definition of the stress tensor by
differentiation of the generating functional, Eq.(\ref{tn-def}).
There is basically no short cut to this problem if one really
needs to determine the whole expression of $\TTT$ including 
the scheme-dependent terms.

An important simplification occurs in deriving
the scheme-independent $\dt=6$ amplitudes: 
firstly, one can limit the inversion (\ref{3p-inv}) to the
15-dimensional subset of $\dt=6$ basic vectors ${\cal P}_i$, $i=1,\dots,15$
(see Table \ref{t_base});  
secondly, these amplitudes are independent of renormalization 
scheme and $T_{\mu\nu}$ definition\footnote{
An argument for the definition-independence is given in Section 5.},
so that eventual constraints among the $u_i$'s,
should be automatically verified in any calculation\footnote{
Note that the null vector has $\dt=4$ and is automatically excluded from 
this basis.}.
More precisely, for $k^2\neq 0$ the
number of amplitudes is 14+2 (transverse+non-transverse)
(see Table \ref{t_sol}):
the two non-transverse tensors are not linearly 
independent in the $\dt=6$ basis but otherwise known,
thus there are 15 unknowns and no constraints.
For $k^2=0$, there are 12+2 unknown amplitudes (the inhomogeneous
ones are still dependent)
and 2 constraints; finally, at the critical point
there are 9+1 unknowns and 5 constraints in both cases. 

Among the $\dt=6$ amplitudes, we are particularly interested
to $A_E$ and $A_W$, which reduce to the anomaly
coefficients $a,c$ at criticality (see Section 3.1).
Hereafter, we present the determination of the $\dt=6$ amplitudes
at the point $k^2=0$, which is sufficient for our later 
purposes (the general case can be found in the Mathematica
notebook associated to this paper). 
The 12+1 independent $k^2=0$ amplitudes are obtained by inverting the
(numeric) 13x15 matrix $Q^\o_i(q^2,k^2=0)$ of Eq.(\ref{3p-inv}) 
restricted to the subspace of ($\dt=6$) ${\cal P}_i$'s.
We remark that this linear problem can be equivalently solved  by
introducing a dual basis of tensors ${\cal T}^*_\o$,
$\left({\cal T}^*_\o , {\cal T}_{\o'}\right) =\delta_{\o,\o'}$,
where the brackets denote a suitable degenerate scalar product
which vanishes on the $\dt <6$ basis \cite{cdgm}.

We finally obtain the following pair of {\it reduction formulae} for
the anomaly amplitudes ($k^2=0$):
\bea
A_E(q^2,0) &=& \frac{1}{36}\left(4 u_{2} - 4 u_{4} - u_{5}
+ 2 u_{8} - 2 u_{11} + 12 u_{14} + 3 u_{15} \right)\ ,
\label{a-redu}
\\
A_W(q^2,0) &=& \frac{1}{108}\left(-8 u_{2} + 8 u_{4} +2 u_{5}
-4 u_{8} +4 u_{11} + 12 u_{14} + 3 u_{15} \right)\ ,
\label{c-redu}
\eea
together with the two constraints:
\beq
u_1 = \frac{1}{2}\ u_3 \ ,\qquad\qquad
u_{12} = 0\ .
\label{u-con}
\eeq
Note that the formulae (\ref{a-redu},\ref{c-redu}) are invariant under
the changes of basis,
${\cal T}_\o \ \to\ {\cal T}_\o + \sum_{i=1}^7 r_i {\cal T}_i$,
$\o=E,W$, which add traceless-transverse terms to the anomaly tensors.

Another interesting reduction formula yields a combination of the 
$\bra TT\ket$ amplitudes which is associated to a $\dt=6$ $\TTT$ 
tensor (see Table \ref{t_sol}):
\beq
{1\over 2 q^2}\left(2 f_2(q^2) +f_0(q^2) -f_0(k^2=0) \right)
 = \frac{u_3}{4}\ ;
\label{3p-f2}
\eeq
this result gives a useful check. 
The complete set of 13 reduction formulae is reported in Appendix D.

At the critical point, the number of independent tensor reduces to
10 and there are 5 constraints; in this case, the reduction formulae
determine the anomaly coefficients (see Eq.(\ref{EW-lim})):
\bea
a &=& \frac{q^2}{12\ \l}\left(
-u_{11} - u_{13} + 4 u_{14} + u_{15} \right)\ ,
\label{a-redu-c}
\\
c &=& \frac{q^2}{36\ \l}\left(
2 u_{11} + 2 u_{13} + 4 u_{14} + u_{15} \right)\ .
\label{c-redu-c}
\eea
The constraints at criticality are:
\bea
u_1 &=& \frac{1}{2}\ u_3 \ ,\qquad\qquad u_{12}\ =\ 0\ ,
\nl
u_2 &=& \frac{1}{4}\left(
4 u_4 + u_{5} - 2 u_{8} - u_{11} -3 u_{13} \right)\ ,
\nl
u_3 &=& \frac{1}{9}\left(
9 u_6 - 2 u_{11} - 2 u_{13} - 4 u_{14} - u_{15} \right)\ ,
\nl
u_9 &=& \frac{1}{4}\left(
- u_{10} + u_{11} + u_{13} \right)\ .
\label{u-con-c}
\eea
Equations (\ref{a-redu-c},\ref{c-redu-c}) allow to compute
 the trace anomalies by standard flat-space perturbative
calculations, and can be useful for interacting theories. 

\noindent
{\bf Infrared divergences}.
As emphasized, the amplitudes $u_i(q^2,k^2=0), i=1,\dots,15$ are ultraviolet
finite and satisfy un-subtracted dispersion relations of the type
(\ref{int-a3})) away from critical points.
However, some amplitudes may develop infrared singularities 
at criticality, which are introduced by the kinematical limit $k^2=0$.
This fact is illustrated by the formula (\ref{3p-f2}) 
(see also Table \ref{t_sol}).
For $k^2\neq 0$, the ($\dt=6$) tensor ${\cal T}_{2-}$
would identify a finite amplitude with safe critical
limit $m\to 0$, namely $f^c_{2-}/q^2\sim \log(q^2/k^2)/q^2$.
For $k^2= 0$, the critical limit  is instead
$f^c_2(q^2)/q^2\sim \log(q^2/m^2)/q^2$, with the physical mass $m$
acting as infrared regulator (rather then the renormalization scale
$\mu$ controlling the ultra-violet singularity
of the $\bra TT\ket$ amplitude (\ref{f2})).

The standard solution to this problem amounts to the choice of
another point $k^2\neq 0$ for the reduction formulae at criticality.
However, this is not necessary for the derivation of the
anomaly amplitudes
Eqs.(\ref{a-redu},\ref{c-redu},\ref{a-redu-c},\ref{c-redu-c}),
which are free of infrared singularities.
Actually, these amplitudes are related to the
trace of the stress tensor, which vanishes as an operator
at criticality due to Weyl invariance;
thus, these amplitudes contain the anomalous pole only.


\subsection{Sum rules for the renormalization-group flow of $a$ and $c$}

We start by recalling the  sum rule 
for the two-dimensional central charge $c$, Eq.(\ref{2d-sum}) \cite{sumrule}:
we present a simplified version of the dispersive proof
of Ref.\cite{cfl}, which can be conveniently generalized to four dimension.
In Section 2.3, we wrote the general form of $\bra TT \ket$ satisfying
the Diff Ward identity, Eq.(\ref{2p-tt}); in two dimensions, 
this reduces to:
\beq
\bra \left(h_1.T(p)\right) \left(h_2.T(-p)\right)\ket  
= f\left(p^2 \right)\ 
    \left(p^2 (h_1) - (p.h_1.p) \right) 
    \left(p^2 (h_2) - (p.h_2.p) \right) \ .
\label{2p-2d}
\eeq
The spin-zero scalar amplitude $f(p^2)$ has dimension $(-2)$ and is
finite and scheme-independent, namely its renormalization is analogous
to that of the four-dimensional amplitudes $A_3(q^2,k^2)$ of 
$\bra AVV\ket$ (Section 2.1) and $A_E, A_W$ of $\TTT$;
note, however, the difference in the single-variable dependence.
The critical limit is:
\beq
f(p^2) \ \to\ - \frac{c\ \pi}{3}\  \frac{1}{p^2}\ ,\qquad\qquad 
\qquad\qquad ({\rm critical\ points}),
\label{f-crit}
\eeq
where $c$ is the Virasoro central charge and $T$ has the standard
two-dimensional normalization  \cite{cfl}.
The amplitude satisfies the un-subtracted dispersion relation:
\bea
f(p^2) &=& \int \ \frac{ds}{\pi} \ {\Im f(s) \over s - p^2 }\ ,
\label{f-disp}\\
\Im f(s) &=& \frac{\pi^2}{3} \ c(s)\ , 
\label{c-id}
\eea
which also identifies $\Im f(s)$ with the 
spectral measure $c(s)$ introduced in (\ref{2d-sum}). 
Its critical limit is:
\beq
c(s)\ \to \ c\ \delta(s)  \ , \qquad\qquad\qquad ({\rm critical\ points}).
\label{c-crit}
\eeq

We now introduce the following sum rule off-criticality,
\beq
\Sigma(m^2) =  \frac{3}{\pi^2} \int_0^\infty \ ds \ \Im f(s,m^2) \ =
\int_0^\infty \ ds \ c(s,m^2)\ ,
\label{2d-sigma}
\eeq
and explicitly write the dependence on the mass scale $m$ introduced by
the relevant (renormalized) coupling constant $g$.
In the following, we would like to prove that
$\Sigma(m^2)$ is actually independent of $m$, 
\beq
\Sigma(m^2) = \Sigma(0)\ ,\qquad \qquad {\rm for}\ \ 0\le m^2 <\infty\ .
\label{m-ind}
\eeq
From Eq.(\ref{c-crit}), we know that $\Sigma(0)= c_{UV}$, 
the central charge of the ultra-violet fixed point.
If $\Sigma$ is independent of $m$, it is also 
renormalization-group invariant, because $t\sim\log(m)$ is the parameter 
of the renormalization-group flow between the fixed points
(UV) and (IR).

Equation (\ref{m-ind}) can be derived from the general property 
that the (trace) anomaly is at most a constant; this was shown in the case
of $\bra AVV \ket$ by using dispersion relations
(see Eqs.(\ref{im-v-cons}-\ref{a-const})): since the imaginary parts of
amplitudes satisfy non-anomalous Weyl Ward identities, the
anomaly should be accounted for by the subtraction constants.
However, the precise formulation of this argument for $\bra TT\ket$
(and later on for $\TTT$) requires the dispersive analysis of all
the amplitudes related by the Ward identity; we prefer to use another
argument based on the dimensional analysis and the renormalization group,
which is presented hereafter.

The amplitude $f(p^2,m^2)$ in (\ref{2p-2d}) is finite and scheme-independent:
there is no wave-function
renormalization because the stress-tensor has no anomalous dimension,
while the coupling-constant renormalization makes the quantity finite
and brings in the scale $m$.
Infrared divergences are also excluded for this quantity in two
dimensions \cite{cfl}.
The imaginary part of $f(p^2,m^2)$ has support on the real positive $s$-axis
and goes to infinity as $o(1/s)$, i.e. faster than the behavior allowed by
its scale dimension, such that the integral in (\ref{2d-sigma}) is finite.
This extra convergence can be verified perturbatively close to the
UV fixed point \cite{cth}\cite{cfl}: 
since $f(p^2,m^2)$ measures the trace anomaly,
it corresponds to an expectation value involving
one $\Theta$ operator (at least); we thus can write,
\bea
f(p^2,m^2) & =& \bra \Theta(p) \cdots \ket 
= \beta(g)\ \bra \phi(p) \cdots \ket \nl
& \sim &
y \ g\ m^y\ s^{-1-y/2}\ ,\qquad\qquad\qquad {\rm for} \ s\to\infty\ ,
\label{large-s}
\eea
where $y >0$ is the dimension of the coupling constant 
($g_o\equiv g\ m^y$), and we used
the relation $\Theta=\beta (g)\ \phi$ expressing the trace off-criticality
as the beta function times the perturbing field \cite{polch}.

In conclusion, $\Sigma(m^2)$ (\ref{2d-sigma}) 
does not contain any regulator for the integrand or the integral;
it depends on a single dimensionful parameter, i.e. $m$,
and thus is necessarily constant (and renormalization-group invariant).

The sum rule (\ref{2d-sigma}) implies some
conditions on its integrand $c(s)$; we can write,
in general \cite{cfl}:
\beq
c(s) = c_0 \ \delta(s) + c_1 (s, m^2) \ .
\label{spec-exp}
\eeq
The function $c_1(s, m^2)$ should be a representation of 
the delta-function with spreading parameterized by $m^2$,
so that the ultra-violet limit $m\to 0$ matches the critical 
form (\ref{c-crit}).
In the opposite limit $m\to\infty$,
$c_1(s, m^2)$ becomes very broad and it contributes
infinitesimally to the correlator $\bra TT \ket$ for any fixed scale $s=s_0$;
therefore, $c(s)$ must be redefined in the infrared scale-invariant 
theory ($m=\infty$) by means of the weak limit:
\beq
c(s) \ \to \ c_0 \ \delta(s)\ , \qquad c_0=c_{IR}\ , \qquad\qquad
(m=\infty).
\label{spec-ir}
\eeq
This Equation identifies $c_0$ with the infra-red value of the central charge
($c_0$ is another renormalization-group invariant).
We can finally rewrite the sum rule (\ref{2d-sigma}) 
in the more familiar form (\ref{2d-sum}) \cite{sumrule}:
\beq
c_{UV} =\int_0^\eps ds\ c(s) \ +\ \int_\eps^\infty ds\ c(s) =
\ c_{IR}\ +\ \int_\eps^\infty\ ds\ c_1(s,m^2) \ .
\label{2d-sum1}
\eeq

This equation implies the $c$-theorem in the integrated form,
$c_{UV} > c_{IR}$, because
the spectral measure $c(s)$ is positive definite \cite{sumrule}.
Furthermore, the integral of $c(s)$ up to a fixed (finite) scale 
defines a function $c(g)$ off-criticality  \cite{cfl}, 
which interpolates between $c_{UV}$ and $c_{IR}$
and satisfies the $c$-theorem in the differential form \cite{cth}:
$dc(g)/d\log(m) \equiv - \beta(g)\ dc(g)/dg <0$.

\bigskip

The four dimensional sum rules for the anomaly coefficients
$a$ and $c$ are obtained by similar arguments.
The corresponding scalar amplitudes $A_\o(q^2,k^2=0)$, $\o=E,W$, 
are extracted from the three-point function $\TTT$ by the
procedure explained in the previous Section; this also applies to the
imaginary parts w.r.t. $q^2$
because the tensors ${\cal T}_\o$ are polynomial at $k^2=0$:
\beq
\Im \TTT = \Im A_E(s,0) \ {\cal T}_E \ +
         \ \Im A_W(s,0) \ {\cal T}_W \ + \cdots\ ,
\qquad\qquad (k^2=0).
\label{3t-im}
\eeq
The amplitudes have dimension $(-2)$ and satisfy 
un-subtracted dispersion relations; their critical limits (\ref{EW-lim})
imply:
\bea
\Im A_E(s,0) & \to & -\pi\l \ a\ \delta(s)\ ,\nl
\Im A_W(s,0) & \to & -\pi\l \ c\ \delta(s)\ , \qquad\qquad
({\rm critical\ points}).
\label{ac-crit}
\eea

The sum rule for the $a$ anomaly is defined in analogy with (\ref{2d-sigma}):
\beq
\Sigma_E\left(m^2,0\right) = - \frac{1}{\pi\l} 
\int_0^\infty\ ds\ \Im A_E \left(s,k^2=0,m^2 \right)\ ,
\label{a-sum}
\eeq
with the $m$-dependence explicitly written. 
This integral is meant to extend over the support of $\Im A_E$.

We now follow the same steps as for the two-dimensional case, 
Eqs.(\ref{m-ind}-\ref{2d-sum1}):
we first argue that $\Sigma_E$ is $m$-independent and equal to the
ultra-violet value $a_{UV}$, provided that it is finite and does not
involve any regulator. 
The integrand is indeed finite and scheme-dependent, as already stressed;
the integral is convergent if the stress tensor is ``improved''
at both the UV and IR fixed points, i.e. $\Theta\equiv 0$ at criticality
(apart from anomalies).
Actually, $\Im A_E$ is related to an expectation value involving
the $\Theta$ operator, whose vanishing at the UV (IR) critical point
forces the asymptotic behavior of $\Im A_E(s,0,m^2)$
to be faster than $1/s$, the scale-invariant law, for $s\to\infty$ 
(respectively, $s\to 0$).
 
The improvement is implied by Weyl invariance, assumed throughout 
this paper, but is not guaranteed for a generic fixed point: as already said,
four-dimensional scale invariance only require $\Theta$ to be
a derivative operator.
Near a generic UV fixed point, it is rather natural to assume the improvement
and the relation $\Theta=\beta(g)\ \phi$,  Eq.(\ref{large-s}) 
\cite{polch}.
Near the IR point, it is less obvious: for example,
the spontaneous symmetry breaking of a global symmetry may
lead to an infrared critical theory without improved stress tensor \cite{cfl}.
We are aware that the hypothesis of improvement at the IR fixed point
(convergence of the sum rule) may exclude interesting 
non-perturbative renormalization-group flows from our analysis.

The constancy of the sum rule (\ref{a-sum}) again constrains the amplitude
to be of the form (\ref{spec-exp}): 
\beq
-\frac{1}{\pi\l}\ \Im A_E \left(s,k^2=0,m^2 \right) =
a_0 \delta(s) + a_1 (s,m^2) \ ;
\label{ae-exp}\eeq
where the smooth function $a_1$ is a
representation of the delta function for $m\to 0$, and the explicit
flow-invariant delta term survives in the infrared theory.
Note that the scale-invariant power law,
$\Im A_E(s,0,0)ds\sim ds/s$, 
is excluded at criticality because it would imply 
a non-vanishing $\Theta$ operator and a divergent sum rule,
as emphasized.

We finally obtain the sum rule for the four-dimensional 
anomaly, analog of Eq. (\ref{2d-sum}):
\beq
a_{UV} - a_{IR}\ = - \frac{1}{\pi\l}\ 
\int_\eps^\infty\ ds\ \Im A_E(s,0,m^2) \ .
\label{wb-a-th}
\eeq
The other four-dimensional anomaly number $c$ satisfies a 
similar sum rule involving $\Im A_W(s,0,m^2)$.
Equation (\ref{wb-a-th}) is the announced result of this Section,
which is followed by a number of remarks: 
\begin{itemize}
\item
If $\Im A_E(s,0,m^2)$ were positive definite,
then Eq.(\ref{wb-a-th}) would prove the $c$-theorem in four dimensions,
$a_{UV} > a_{IR}$,
as suggested by the non-trivial examples mentioned in the Introduction
\cite{afgj}\cite{fgpw}.
We do not discuss the positivity in this paper, but only add 
some remarks in the Conclusions. 
\item
On the other hand, the corresponding result for the Weyl-squared term, 
$c_{UV} > c_{IR}$, is not true in general, as shown by known
counter-examples \cite{cfl}\cite{afgj};
the respective amplitude $\Im A_W(s,0,m^2)$ cannot be positive definite 
in general, although its critical limit {\it is} always positive, 
i.e. $c_{UV},c_{IR}>0$, due to the positivity of $\bra T(x)T(0)\ket$
at criticality ($|x|\neq 0$) \cite{cfl}.
These facts suggest the absence of naive back-of-the-envelope 
positivity arguments.
\item
The other three $\dt=6$ amplitudes $A_i$
corresponding to traceful tensors ${\cal T}_i$
($i=15,\dots,17$ in Table \ref{t_sol})
also satisfy analogous sum rules; however, their vanishing critical
limit imply null sum rules (\ref{a-sum}), $\Sigma_i=0$, for $i=15,\dots,17$: 
thus, the imaginary parts of these amplitudes cannot have definite sign.
\item
We stress that the positivity of $\Im A_E(s,0,m^2)$ is not
necessary for the derivation of the sum rule (\ref{wb-a-th});
its smooth part $a_1(s,m^2)$
can be any representation of the delta function which becomes
very broad in the infrared limit $m\to\infty$ (this
is assured by the decoupling of massive states).
\end{itemize}

We conclude this section with a remark on the chiral anomaly,
analysed in Section 2.1:
the sum rule (\ref{wb-a-th}) can also be written
for the flow of the chiral anomaly, as is clear from Eq.(\ref{a-const}),
analogous to (\ref{a-sum}). 
However, there is an important difference:
the role of the Weyl symmetry is now played by the chiral symmetry,
such that ``critical theory'' should be replaced with 
``theory with massless fermions''. 
These are not equivalent properties, because it is conceivable a 
renormalization-group flow with chiral symmetry, for example a QCD-like
theory with massless quarks. In this case, the relevant operator
causing the flow
yields the stress-tensor trace according to (\ref{large-s}), but does not
couple to the chiral current; as a consequence, 
$\Im A_3(s,0)$ in (\ref{a-const}) remains equal to its UV critical 
form of a delta function and there is no actual flow of the
chiral anomaly \cite{fsby}:
\beq
C_{UV}=C_{IR} \ ,\qquad\qquad\qquad{\rm (RG\ flows\ in\ chiral\ theories).}
\label{thooft}\eeq
Furthermore, the term $\Im \bra AVV\ket \sim\delta(s)$ 
means that the chiral anomaly
is saturated by massless on-shell intermediate states ($q^2=k^2=0$)
\cite{cole}; thus, it can be perturbatively computed either from 
the fundamental (UV) or the composite (IR) massless fermions;
Equation (\ref{thooft}) corresponds to the `t Hooft anomaly matching 
conditions \cite{fsby}.
In conclusion, the sum rules for the conformal and chiral anomalies are
mathematically identical but may lead to rather different physical 
consequences.

\setcounter{footnote}{1}

 
\section{Test calculations in free field theories}
 
We now apply the results of Section 4 to the cases of the free massive scalar
and Dirac fermion theories\footnote{Background material for these
calculations can be found e.g. in Ref.\cite{cfl}.};
we (re)compute here the anomaly coefficients $a, c$,
verify the related sum-rules and check other properties
of the reduction formulae.

The stress-tensors three-point correlator is obtained, 
according Eq.(\ref{tn-def}), by triple variation  w.r.t $g^{\mu \nu} $ of 
the generating functional:
\beq
W(g)\equiv \log \int [d\phi] e^{I(\phi;g)} \ .
\eeq 
After variations and   flat limit   one schematically obtains:
\bea
\bra T_1 T_2 T_3\ket
&=&
\bra  \delta_1 I \delta_2 I\delta_2 I 
      +\left(\delta_1 I \delta_2\delta_3 I + perms. \right) 
      +\delta_1 \delta_2\delta_3I\ket
+\cdots,
\label{tttdiags}\eea
where  we denoted by $perms.$ the permutations of labels and
we left unspecified some disconnected terms, all containing a 
$\bra \delta I\ket $ factor.
In terms of Feynman diagrams, the first 
three contributions to $\TTT$ in  Eq. (\ref{tttdiags}) 
are described respectively by three one graviton vertices
simply linked in a triangle, by
a two graviton vertex doubly linked to a one graviton vertex, and
a tadpole over a three graviton vertex.

It is clear that a complete calculation of all contributions 
-- even in a free field theory -- is a nontrivial algebraic task.
Nevertheless, it should be reminded that the results of Section
4 only involve the subset of $\TTT$ terms that have 
tensorial dimension $\dt=6$.
We now remark the following simple property: if 
the action $I$ contains at most two derivatives, any insertions of
(variations of) $I$ in the expectation value can raise the tensorial
dimension by two at most. It then follows that the
second and third contributions to $\TTT$ have tensorial dimension
$\dt <6$ and should not be considered for our purposes.
Moreover, owing to the fact that 
$\bra \delta_{g^{\mu \nu}} I\ket \propto \delta_{\mu \nu}$ in a flat space, 
the leftover disconnected terms in (\ref{tttdiags}) 
also have $\dt <6$.
 
In conclusion, we are left  with  the triangle diagram whose explicit
contribution is,  
\bea
\TTT_{{\rm triangle}}&=&
\int \!\!{d^4L\over (2 \pi)^4}
V_{h_3}(L-q,-L)S(L) V_{h_1}(L,k_1-L)S(L-k_1) \nl 
&&\qquad\qquad\times
V_{h_2}(L-k_1,q-L)S(L-q)\ ,
\eea
where the vertex $V(k_1,k_1)$ and the propagator $S(p)$
can be found, e.g.  in Ref.\cite{cfl}(see Equations (3.27-3.30) there).

The imaginary part w.r.t. $q^2$ of $\TTT_{{\rm triangle}}$ 
for $q^2>0$ and $k^2=0$
is  obtained by the standard Cutkowsky rules, cutting the lines 
linked to the vertex $T_3$ (the remaining integrals  are elementary).
After collecting from the resulting expression
the coefficients $u_i(q^2,0)$ of the expansion in the canonical $(\dt=6)$ 
basis ${\cal P}_i$, $i=1,\dots15$, we have tested the formulas of Section 4.
We have found that:
\begin{itemize}
\item 
The constraints (\ref{u-con}) and (\ref{u-con-c}) are satisfied 
off and at criticality, respectively.
\item
The reduction formulae (\ref{a-redu},\ref{c-redu}) determine
the following expressions for the Euler and Weyl anomaly amplitudes,
in the massive scalar theory:
\end{itemize}
\bea
&-& {1\over \lambda\pi} \Im A_E(s,0,m^2) \nl
\bh\bh\bh &=& 
\theta(x-1)\ \frac{5}{96 m^2 x^4} \left[
32 r x (1+3x) +(-27 +80 x +12 x^2)\log\left(\frac{1-r}{1+r}\right)
\right] ,
\label{bose-a}\\
&-& {1\over \lambda\pi} \Im A_W(s,0,m^2) \nl
\bh\bh\bh &=& 
- \theta(x-1)\ \frac{5}{144 m^2 x^4} \left[
8 r x (13+3x) +(27 +26 x +12 x^2)\log\left(\frac{1-r}{1+r}\right)
\right] ,
\label{bose-c}
\eea
with
\beq
x= \frac{s}{4m^2}\ ,\qquad\qquad r=\sqrt{1-\frac{4m^2}{s}}\ .
\eeq
In the massive fermion theory, we find:
\bea
&-& {1\over \lambda\pi} \Im A_E(s,0,m^2) \nl
\bh\bh\bh &=&
- \theta(x-1)\ \frac{5}{24 m^2 x^4} \left[
8 r x (4+17x) +3(-9 +25 x +12 x^2)\log\left(\frac{1-r}{1+r}\right)
\right] ,
\label{fermi-a}\\
&-& {1\over \lambda\pi} \Im A_W(s,0,m^2) \nl
\bh\bh\bh &=&
\theta(x-1)\  \frac{5}{72 m^2 x^4} \left[
4 r x (52+5x) +3(18 +23 x -3 x^2)\log\left(\frac{1-r}{1+r}\right)
\right] .
\label{fermi-c}
\eea
\begin{itemize}
\item
The sum rules (\ref{wb-a-th}) for these imaginary amplitudes
verify the expected renormalization-group flows of the trace anomalies  
(see e.g. Ref.\cite{cfl} for the well-known free-field anomalies):
\bea
a_{UV}-a_{IR} &=& 1\ , \qquad c_{UV}-c_{IR} = 1\ , \qquad\qquad\qquad
(\rm scalar);
\nl
 a_{UV}-a_{IR} &=& 11\ , \qquad c_{UV}-c_{IR} = 6\ , \qquad\qquad
(\rm Dirac\ fermion).
\eea
\item
The amplitudes $A_i$, $i=15,16,17,$ of the traceful tensors vanishing at 
criticality have also been computed (using the formulas in Appendix D); 
the corresponding sum rules (\ref{a-sum})
are convergent and equal to zero, as expected:
\beq
\Sigma_i =0 \ , \qquad\qquad i=15,16,17.
\eeq
\item
The imaginary parts of the Euler and Weyl amplitudes turn out to be 
positive definite in all the above examples but one, i.e. the bosonic
$\Im A_E$ (\ref{bose-a}); nevertheless, a positive
definite function can be obtained by adding to
$\Im A_E$ a suitable linear combination of the three non-definite
amplitudes $\Im A_i$, $i=15,16,17,$ which have vanishing sum rule.
\item 
Infrared divergences occur at criticality
for some of the amplitudes $A_i$, $i=1,\dots,7$, of the traceless tensors,
in agreement with the discussion in Section 4.1; moreover, the formula 
$f_2(q^2)/q^2\sim \log(q^2/m^2)/q^2$ is checked for $m\to 0$
(including the numerical coefficient computing $c$ again).
\end{itemize}
We finally remark that another possible test of the reduction
formulae at criticality (\ref{a-redu-c},\ref{c-redu-c},\ref{u-con-c})
is provided by the free massless vector field;
however, the theory becomes interacting off-criticality 
(there is no flow into the massive free vector), so that
the sum rules can only be checked perturbatively.
We postpone this analysis to further developments of our work.

 
\section{Conclusions} 

In this paper, we have found the general four-dimensional form 
of the stress-tensor three-point function, written as an expansion
over scalar amplitudes times tensor structures solution
of the Ward identities. We thus made explicit the conditions imposed
by the mandatory symmetries, leaving out any dynamical
hypothesis. From this result, we have identified the amplitudes
corresponding to the two universal trace-anomaly numbers $a$ and $c$, and
derived the renormalization-group sum rules which they satisfy.

The main motivation for this analysis has been the extension 
of the Zamolodchikov $c$-theorem to four dimensions:
we did not provide a proof but we tried to clear the ground
for further attempts to it. 
One possible road could be the proof of positivity 
for the imaginary amplitude entering the
$a$ sum rule (\ref{wb-a-th}) (possibly using the weak-energy condition
of Ref.\cite{ol}). 

Other arguments which stands on specific dynamical 
hypotheses can hopefully take advantage of our reduced set of variables.   
For example, exploiting the conditions of ${\cal N}=1$ supersymmetry,
which is present in most of the non-trivial tests of the would-be 
$a$-theorem \cite{afgj}\cite{fgpw}.
It would also be interesting to reformulate the dynamical
assumptions of Refs.\cite{ath} at the level of three-point functions. 
Finally, the analysis of manifestly positive-definite four-point amplitudes
is feasible with the technology presented in this paper.

\bigskip

{ \bf Acknowledgements}  \\
The authors thank the hospitality of each other authors' Institution
at various stages of this collaboration.
They also thank D. Anselmi, D. Atlani, C. M. Becchi, J. Bros 
and Y. Stanev for useful discussions.
This work was supported in part by the European Community  
Network grant FMRX-CT96-0012.

\appendix

 
\section{Notations and conventions} 
\label{app-conv}

The four-dimensional trace anomaly is written:
\beq
 g^{\mu\nu} \bra T_{\mu\nu} \ket \equiv 
\bra \Theta \ket = \lambda \left(
a\ E -3c\ W +a'\ D^2 {\cal R}+ r\ {\cal R}^2 \right)\ , \qquad
\lambda\equiv \frac{1}{2880\cdot 4\pi^2}\ .
\label{an-app}\eeq 
where the Euler density and the square of the Weyl tensor are,
respectively,
\bea
E&\equiv &\left( R_{\mu_1\mu_2\mu_3\mu_4}\right)^2 
 -4 \left({\cal R}_{\mu_1\mu_2}\right)^2 +{\cal R}^2\ ,
\nl
W&\equiv& \left(C_{\mu_1\mu_2\mu_3\mu_4}\right)^2 = 
\left(R_{\mu_1\mu_2\mu_3\mu_4} \right)^2 
-2\left({\cal R}_{\mu_1\mu_2} \right)^2
+{1\over 3} {\cal R}^2\ ,
\eea
and $D^2$ is the square of the covariant derivative.
The Riemann and Ricci tensors are:
\bea \label{rie} 
R^{\mu_1}_{\ \mu_2 \mu_3 \mu_4} &\equiv&
\de_{\mu_3} \Gamma^{\mu_1}_{\mu_4 \mu_2}-\de_{\mu_4} 
\Gamma^{\mu_1}_{\mu_3 \mu_2} 
+\Gamma^{\mu_1}_{\mu_3 \mu_5}\Gamma^{\mu_5}_{\mu_4 \mu_2} 
-\Gamma^{\mu_1}_{\mu_4 \mu_5}\Gamma^{\mu_5}_{\mu_3 \mu_2}\ ,
\\
\label{ricci}
{\cal R}_{\mu_2 \mu_4}&\equiv& R^{\mu_1}_{\mu_2\mu_1\mu_4}\ ,
\qquad\qquad {\cal R} \equiv {\cal R}^\mu_\mu\ .
\eea 
The LeviCivita connection is:
\beq \label{levi} 
\Gamma^{\mu_1}_{\mu_2 \mu_3}\equiv \half g^{\mu_1 \mu_4}
\left(\de_{\mu_2} g_{\mu_4 \mu_3}+\de_{\mu_3} 
g_{\mu_4 \mu_2}-\de_{\mu_4} g_{\mu_2 \mu_3} \right) \ ,
\eeq 
and the metric has Euclidean signature.
We also use the Minkowski signature $(+,---)$ when discussing 
dispersion relations (Sections 2.1 and 4.2); on this metric,
the trace anomaly (\ref{an-app}) changes sign and the normalization 
$\l$ should be chosen negative.

For any local quantity ${\cal A}(x;g)$ functional of $g^{\mu \mup}(x)$,
we consider its $n$-th variation at the point 
$g^{\mu \mup }=\gflat^{\mu \mup }$ and contract the pairs of indices
with generic polarization tensors $h_i^{\mu\mup}$ as explained at
the beginning of Section 2.3. We denote the $n$-th variation by:
\beq \label{var-def} 
\bh {\cal A}\left(x_1,h_1;\dots; x_n,h_n;x \right)
\equiv 
h_1^{\mu_1\mup_1}{\delta\over\delta g^{\mu_1\mup_1}(x_1)}
\cdots 
h_n^{\mu_n\mup_n}{\delta\over\delta g^{\mu_n\mup_n}(x_n)}
{\cal A}(x;g) \ .
\eeq
The differential of a functional $F[g]$ w.r.t. the metric, 
$g^{\mu\mup}\to g^{\mu\mup}+ \l^{\mu\mup}$, with $\l$ an infinitesimal
function, is:
\beq
\delta_\l\ F[g] = F[g+\l] -F[g]=\int dx 
{\delta F\over\delta g^{\mu\mup}(x)}\ \l^{\mu\mup}(x)\ .
\eeq

The Fourier transform of translation-invariant functions 
$f(x_1,x_2,..,x_n)$ is defined by:
\beq\label{four-def} 
f(x_1,x_2,..,x_n)=(2\pi)^{d-nd }\int dp_1...dp_n \ 
\delta\left(\sum_{i=1}^n p_i\right)\ 
\exp\left(i\sum_{i=1}^n p_i\cdot x_i\right)\   f(p_1,......,p_n) \ .
\eeq
In particular,  the metric variation
(\ref{var-def}) is translation invariant after flat-space limit 
 and its Fourier transform 
is denoted by ${\cal A}\left(k_1,h_1;\dots;k_n,h_n\right)$.


\section{Diff and Weyl Ward identities} 
\label{app-ward}

The Diff Ward identities for the stress-tensor $n$-point functions
are obtained by performing the reparametrization 
$\d_\eps$ on the argument of the generating functional
$W[g+h]$; here, $g^{\mu\nu}$ is the background metric and 
$h^{\mu\nu}$ is a small deformation, which is used as a source
for extracting the stress-tensor correlators according to (\ref{tn-def}).
The Diff transformations are:
\bea
\delta_\eps g^{\mu\nu} &=& D^{(\mu} \eps^{\nu)}\ ,
\label{lie-d1}\\
\delta_\eps h^{\mu\nu} &=& \left( D_\s \eps^{(\mu} \right)
h^{\s\nu)} - \left( \eps\cdot D\right) h^{\mu\nu}\ .
\label{lie-d2} 
\eea
Upon expanding $W[g+h]$ in series of $h$ and performing the
transformations, one finds that each term $O(h^n)$, $n=0,1,\dots$ yields
the Ward identity for the $(n+1)$-point function (also involving
the $n$-point correlator); the local form of the
identity is obtained after integration by parts of 
the derivatives in (\ref{lie-d1},\ref{lie-d2})
and the limit to flat space.
To order $O(h^0)$, one finds $\partial_\mu \bra T_{\mu\nu} \ket =0$,
and to order $O(h)$ the two-point identity Eq.(\ref{2p-difx}).
The three-point identity (\ref{3p-diff}) becomes simpler
in momentum space and after contraction
of indices with polarization tensors, as explained in Section 2.3.

The Weyl Ward identities are similarly obtained from 
Eq. (\ref{ward-def}), which we rewrite here:
\beq
\d_\s W[g] \equiv -2 \int \sqrt{g}\ g^{\mu\nu}\ \bra T_{\mu\nu}\ket\ \sigma 
= -2 \int dx\ \sqrt{g}\  {\cal A}(x;g)\ \sigma(x)\ ,
\label{weyl-def}\eeq
with anomaly given by (\ref{an-app}).
Higher-point functions are obtained by differentiating both sides of 
this Equation w.r.t. to $g^{\mu\nu}$
($\int h \cdot \d/\d g$): on the left hand side, the variation can act on 
the explicit metric used for the trace or on v.e.v. yielding another
stress-tensor; on the
l.h.s., it can act on $\sqrt{g}$ or on the anomaly 
(whose variations are written in the notations 
(\ref{var-def},\ref{four-def})).
After two variations, flat-space limit and Fourier transform, one finds 
Eq.(\ref{3p-weyl}), which we rewrite here:
\bea\label{3p-wey} 
&&\bh\bh \bra \T(k_3)(T(k_1).h_1)(T(k_2).h_2) \ket \nl 
&+& \left[ \bra (T(-k_2).h_1)(T(k_2).h_2) \ket + 
                     \bra (T(k_1).h_1)(T(-k_1).h_2) \ket \right] \nl 
&= & \A (k_1,h_1;k_2,h_2) - \half \left[ 
(h_1) \A (k_2,h_2) + (h_2) \A (k_1,h_1) \right] \ .
\eea 

As said in the text, the compatibility of this identity and its
analogue for $(k_3 \leftrightarrow k_1)$ imply the vanishing of the
${\cal R}^2$ term in the anomaly ($r=0$), in agreement with the Wess-Zumino
consistency condition \cite{chi-book}\cite{cc}.
The compatibility of the Diff and Weyl identities yields
the equation:
\bea 
0&=& {\cal A}(k_1,k_1\otimes v +v\otimes k_1;k_2,h_2) \nl
& +& {\cal A}(k_1+k_2, v\otimes (h_2 . k_1) + (h_2 . k_1) \otimes v) \nl
& -& (k_3.v) {\cal A}(k_2,h_2)  
- (k_2.v) {\cal A}(k_1+k_2,h_2) 
- (k_1.v) {\cal A}(k_2,h_2) \nl
& -& \half (h_2) {\cal A}(k_1,v\otimes k_1 +k_1 \otimes v) \ ,
\label{compat1}
\eea 
which provides an useful check for the variations
of the anomaly. The first variation of the Euler and Weyl terms 
vanishes in flat space and thus their second variation is transverse.

The conformal isometries of the metric,
\beq
0=\left(\d_\eps +\d_\s \right) g^{\mu\nu} (x)\equiv
D^{(\mu} \eps^{\nu)} -2\s(x) g^{\mu\nu}(x)\ ,
\label{lie-d3}\eeq
imply Ward identities which are linear
combinations of the previous Diff and Weyl identities.
Let us consider the integrated expressions of the latter
equations, say the $O(h^n)$ terms: these
contain the $(n+1)$-point function multiplying
$\d_\eps g^{\mu\nu}$ and $\d_\s g^{\mu\nu}$, respectively,
plus the $n$-point functions and the $n$-th variation of the
anomaly for the Weyl identity.
By summing the two identities, the $(n+1)$-point functions cancel out for
conformal isometries (\ref{lie-d3}); after integration by
parts of $\d_\eps h$ in (\ref{lie-d2}) plus $\d_\s h = -2\s h$,
with vanishing surface term \cite{cardy2}, the remaining $n$-point function
contains the ``conformal Lie derivative'' 
$(\d_\eps +\d_\s) T(x_i)$ (\ref{conf-t}) acting on each stress tensor,
$i=1,\dots,n$; finally, there is the term with
$n$-th variation of the trace anomaly. 
The result for $n=2$ is given by Eq.(\ref{2p-conf}).


\section{Construction of the basic six-index tensors} 
\label{app-tens}

In this Section we construct the tensorial basis 
on which the correlators are decomposed. 
 We start by considering the most general basis 
of linearly independent tensors with $2 n$ indices ($n=1,2,3$),  
$\mu_1\mup_1\cdots \mu_n\mup_n$, 
that are symmetric by exchange of any couple $\mu_i \leftrightarrow \mup_i$. 
In addition we require those tensors to be built in terms of the  
available covariant quantities:  
${k_1}_{\mu},{k_2}_{\mu},\gflat_{\mu\mup}$. 
It is evident that all the symmetric  2-tensors made of 
$k_1,k_2,\gflat$ are generated by the following  $\nta{2}=4$~  
{\bf 2-tensors}:   
\beq\label{2-tens} 
 (k_1.h_1.k_1), (k_2.h_1.k_2), (k_1.h_1.k_2), (h_1)  
\eeq 
(indices are contracted with the polarization tensor $h_1$
as explained in Section 2.3). 
  
The  construction of  symmetric $4$ and $6$ tensors made of $k_1,k_2,\gflat$ 
is simplified by introducing the idea of primitive tensor: 
a primitive tensor is a tensor that is made by a connected chain of  
index contractions.  
For instance $(k_1.h_1.h_2.k_2)$ is primitive,  
while $(k_1.h_1.h_2.k_2)(h_3)$ is not. 
All the 2-tensors of Eq.(\ref{2-tens}) are primitive. 

Here follows the  list of the $\ntp{4}=5$ symmetric {\bf primitive 4-tensors}:
\bea 
(k_1.h_2.h_1.k_1),(k_2.h_2.h_1.k_2),(k_1.h_2.h_1.k_2),(k_2.h_2.h_1.k_1), 
(h_1.h_2),
 \eea 
and of the $\ntp{6}=13$ symmetric {\bf primitive 6-tensors},
\bea& &(k_1.h_3.h_1.h_2.k_1),(k_1.h_3.h_2.h_1.k_1),(k_1.h_2.h_3.h_1.k_1),\nl 
& &(k_2.h_3.h_1.h_2.k_2),(k_2.h_3.h_2.h_1.k_2),(k_2.h_2.h_3.h_1.k_2),\nl 
& &(k_2.h_3.h_1.h_2.k_1),(k_2.h_3.h_2.h_1.k_1),(k_2.h_2.h_3.h_1.k_1),\nl 
& &(k_1.h_3.h_1.h_2.k_2),(k_1.h_3.h_2.h_1.k_2),(k_1.h_2.h_3.h_1.k_2),\nl 
& &(h_1.h_2.h_3) 
. 
\eea 
 
The complete basis of symmetric $2n$-tensors can be constructed by  
multiplying in all possible ways primitive  
symmetric $2 k$ tensors ($1\leq k\leq n$)  
in such a way that the total number of free indices is $2n$. 
In particular it is easily checked that the  4-tensors basis has  
dimension $\nta{4}=21=\ntp{4}+\ntp{2}^2$,  
while the 6-tensors are $137= \ntp{6}+ 3\ \ntp{4}\ \ntp{2}+\ntp{2}^3$.   
  
For our purposes we also need  the basis  
of 5-tensors $E_{\mu_1\mup_1 \mu_2 \mup_2 \mu_3}$ made of  
$k_1,k_2,\gflat$, that are
symmetric by exchange of $\mu_i \leftrightarrow \mup_i$; they 
are contracted with $h_1,h_2$ and one polarization vector $v$. 
Again, the complete basis is found by multiplying 
primitive tensors in such a way to obtain 5-tensors. 
In this case, we also need to consider the $\ntp{1}=2$  
{\bf primitive 1-tensors}:
\beq 
 (k_1.v), (k_2.v);
\eeq 
the $\ntp{3}=2$ {\bf primitive 3-tensors}:  
\beq 
(k_1.h_1.v),(k_2.h_1.v);
\eeq   
and the $\ntp{5}=4$ {\bf primitive 5-tensors},
\beq 
(v.h_2.h_1.k_1),(k_1.h_2.h_1.v),(v.h_2.h_1.k_2),(k_2.h_2.h_1.v). 
\eeq   
We easily obtain   
$\nta{5}=62=\ntp{5}+\ntp{4}\ntp{1}+2\ntp{3}\ntp{2}+\ntp{1}\ntp{2}^2$  
5-tensors. 
 
Next, we need to impose the Bose symmetry  
realized by the exchange of $(k_1,h_1)\leftrightarrow (k_2,h_2)$
on the 4-tensors and 6-tensors. This can be done by looking at  
the action on the primitive tensors: it is easy to realize
then that the  
Bose symmetric 4-tensors are $\ntb{4}=14$ and that the  
Bose symmetric 6-tensors are $\ntb{6}=77$.  
  
In Section 2.3, we defined the {\it tensorial dimension}
 $\dt$ of a tensor monomial as being the total number of $k_1,k_2$ with 
free indices  (i.e. contracted with polarization tensors $h_i$ or $v$).  
For instance $\dt ( (k1.h2.k2)(k2.h1.k2))= 4$. 
The tensor dimension of a polynomial is the maximal tensorial 
dimension of the constituent monomials. 
Hereafter we report the dimensions of the tensors basis, 
divided according to the $\dt$ value, writing
$\ndt{m}{\dt}$ for "$m$ tensors of tensorial dimension $\dt$". 
    
Classification of primitive  tensors: 
\bea 
 & &\ntp{1}=2=\ndt{2}{1}\ ,\nl 
 & &\ntp{2}=4=\ndt{3}{2}+\ndt{1}{0}\ ,\nl 
 & &\ntp{3}=2=\ndt{2}{1}\ ,\nl 
 & &\ntp{4}=5=\ndt{4}{2}+\ndt{1}{0}\ ,\nl 
 & &\ntp{5}=4=\ndt{4}{1}\ ,\nl 
 & &\ntp{6}=13=\ndt{12}{2}+\ndt{1}{0}\ . 
\eea  
 
Classification of all tensors: 
\bea 
 & &\nta{2}=4=\ndt{3}{2}+\ndt{1}{0}\ ,\nl 
 & &\nta{4}=21=\ndt{9}{4}+\ndt{10}{2}+\ndt{2}{0}\ ,\nl 
 & &\nta{5}=62=\ndt{18}{5}+\ndt{32}{3}+\ndt{12}{1}\ ,\nl 
 & &\nta{6}=137=\ndt{27}{6}+\ndt{63}{4}+\ndt{42}{2}+\ndt{5}{0}\ . 
\eea   

Classification of all Bose-symmetric tensors: 
\bea 
 & &\ntb{4}=14=\ndt{6}{4}+\ndt{6}{2}+\ndt{2}{0}\ ,\nl 
 & &\ntb{6}=77=\ndt{15}{6}+\ndt{34}{4}+\ndt{24}{2}+\ndt{4}{0}\ . 
\eea


\section{Long formulae}
\label{app-sol}

Here we report the explicit form of the ``null vector''
which occurs  in the $(1\leftrightarrow 2)$ symmetric
six-index tensors basis ${\cal P}_i$, $i=1.\dots,77,$
(Table \ref{t_base}) in four dimensions;
this vector expresses the linear dependence in this basis 
and should be put to zero (Section 3).
The null vector is ${\cal T}_8$ in Eq.(\ref{3p-end}); it reads:
\bea
{\cal T}_8 
&=& 2\ (\p_{17}+\p_{20}-\p_{26}-\p_{27}+\p_{29}
      -\p_{33}-\p_{34}+\p_{46}+\p_{49} ) 
  -\p_{36}-\p_{39}-\p_{41}-\p_{43} \nl
&+& 2 k^2\ (\p_{50}+\p_{51}+\p_{52}-\p_{60}-\p_{61}-\p_{66} ) \nl
&+&  \left( 2k^2 - q^2 \right) 
( \p_{53}+\p_{54}+\p_{55}+\p_{56}+\p_{71}+\p_{73}
-\p_{59}-\p_{62}-\p_{63}-\p_{65}-\p_{67}-\p_{68} ) \nl
&+&  k^2\ (-\p_{57}-\p_{58}-\p_{64}+\p_{69}+\p_{70}+\p_{72} ) \nl
&+&  \frac{q^2}{4}\left( 4k^2 - q^2 \right) \ 
   (-2\p_{74}+\p_{75}+\p_{76}-\p_{77}) = 0 \ .
\label{nul-vec}\eea
Note that this tensor has $\dt=4$, because it possesses
no components in the first 15 elements of the basis;
moreover, it is easy to check that its scale dimension is $\D=4$

\bigskip  

Next, we write the complete set of reduction formulae (Section 4.1) which
determine the 13 $\TTT$ amplitudes $A_i(q^2,k^2=0)$ 
of the $\dt=6$ tensors (see Table \ref{t_sol}): 
for $i=1,\dots,7$, they correspond to the traceless transverse
tensors, for $i=9,10,$ to the anomalies ($A_E,A_W$),
and for $i=15,16,17,$ to the traceful transverse tensor that
vanish at the Weyl-invariant critical points.
The formulae express the amplitudes in terms of the coefficients $u_i$
of the $\dt=6$ tensors ${\cal P}_i$, $i=1,\dots,15$.

Three amplitudes and the two constraints among the $u_i$'s were
already given in the text, see Eqs.(\ref{a-redu}-\ref{3p-f2}); 
the remaining 7+3 amplitudes are:
\bea
A_{1}&=& u_{14}\ ,\qquad A_{2} = u_{2}\ ,\qquad 
A_{4} = u_{4}\ ,\nl
A_{3}&=& \frac{1}{36}
\left( 4u_{2}-27u_{3}-4u_{4}-u_{5}+9u_{7}+11u_{8}-2u_{11}+
       12u_{14}+3u_{15} \right)\ ,\nl
A_{5}&=& \frac{1}{72}
\left(-28u_{2}+81u_{3}+28u_{4}+7u_{5}-14u_{8}+5u_{11}-12u_{14}-
       3u_{15} \right)\ ,\nl
A_{6}&=&\frac{1}{288}
\left(-40u_{2}+99u_{3}+40u_{4}+10u_{5}-20u_{8}-18u_{9}+20u_{11}-
      12u_{14}-3u_{15}\right)\ ,\nl
A_{7}&=& u_{7}\ ,
\eea
and
\bea
A_{15}&=& \frac{1}{3}
\left( 4u_{2}-4u_{4}-u_{5}+2u_{8}+12u_{9}+3u_{10}-2u_{11} \right)\ ,\nl
A_{16}&=& \frac{1}{3}
\left( 4u_{2}-4u_{4}-u_{5}+2u_{8}+u_{11}+3u_{13} \right)\ ,\nl
A_{17}&=& \frac{1}{27}
\left( 8u_{2}-27u_{3}-8u_{4}-2u_{5}+27u_{6}+4u_{8}-4u_{11}-
            12u_{14}-3u_{15} \right)\ .
\eea
The corresponding formulae for non-vanishing $k^2$ can be
found in the Mathematica notebook. 
At criticality, there are 3 further constraints (\ref{u-con-c})
among the $u_i$'s, which are equivalent to the conditions
$A_i=0$, for $i=15,16,17$.


\def\NPB#1#2#3{{\it Nucl.~Phys.} {\bf{B#1}} (#2) #3} 
\def\CMP#1#2#3{{\it Commun.~Math.~Phys.} {\bf{#1}} (#2) #3} 
\def\CQG#1#2#3{{\it Class.~Quantum~Grav.} {\bf{#1}} (#2) #3} 
\def\PLB#1#2#3{{\it Phys.~Lett.} {\bf{B#1}} (#2) #3} 
\def\PRD#1#2#3{{\it Phys.~Rev.} {\bf{D#1}} (#2) #3} 
\def\PRL#1#2#3{{\it Phys.~Rev.~Lett.} {\bf{#1}} (#2) #3} 
\def\ZPC#1#2#3{{\it Z.~Phys.} {\bf C#1} (#2) #3} 
\def\PTP#1#2#3{{\it Prog.~Theor.~Phys.} {\bf#1}  (#2) #3} 
\def\MPLA#1#2#3{{\it Mod.~Phys.~Lett.} {\bf#1} (#2) #3} 
\def\PR#1#2#3{{\it Phys.~Rep.} {\bf#1} (#2) #3} 
\def\AP#1#2#3{{\it Ann.~Phys.} {\bf#1} (#2) #3} 
\def\RMP#1#2#3{{\it Rev.~Mod.~Phys.} {\bf#1} (#2) #3} 
\def\HPA#1#2#3{{\it Helv.~Phys.~Acta} {\bf#1} (#2) #3} 
\def\JETPL#1#2#3{{\it JETP~Lett.} {\bf#1} (#2) #3} 
\def\JHEP#1#2#3{{\it JHEP} {\bf#1} (#2) #3} 
\def\TH#1{{\tt hep-th/#1}}

\end{document}